\documentclass[a4paper,11pt]{article}
\pdfoutput=1 % if your are submitting a pdflatex (i.e. if you have
\usepackage{jheppub} % for details on the use of the package, please
\usepackage[T1]{fontenc}
\usepackage[all]{xy}
\usepackage[dvipsnames]{xcolor}
\usepackage{graphicx}% Include figure files
\usepackage{dcolumn}% Align table columns on decimal point
\usepackage{bm}% bold math
\usepackage{hyperref}% add hypertext capabilities
\usepackage{graphicx}
\usepackage{enumerate}

\newcommand{\slfrac}[2]{\left.#1\middle/#2\right.}

\makeatletter
\DeclareFontFamily{U}{tipa}{}
\DeclareFontShape{U}{tipa}{m}{n}{<->tipa10}{}
\newcommand{\arc@char}{{\usefont{U}{tipa}{m}{n}\symbol{62}}}%

\newcommand{\arc}[1]{\mathpalette\arc@arc{#1}}

\newcommand{\arc@arc}[2]{%
  \sbox0{$\m@th#1#2$}%
  \vbox{
    \hbox{\resizebox{\wd0}{\height}{\arc@char}}
    \nointerlineskip
    \box0
  }%
}
\makeatother

\begin{document}

\preprint{APS/123-QED}

\title{Boundary electromagnetic duality from homological edge modes}% Force line breaks with \\

\author[a]{Philippe Mathieu,}
\author[b]{Nicholas Teh}

\affiliation[a]{Department of Mathematics, University of Notre Dame, Indiana, USA}
\affiliation[b]{Department of Philosophy, University of Notre Dame, Indiana, USA}

\emailAdd{pmathieu@nd.edu}%Lines break automatically or can be forced with \\
\emailAdd{nteh@nd.edu}

\date{\today}% It is always \today, today,

\abstract{
Recent years have seen a renewed interest in using `edge modes' to extend the pre-symplectic structure of gauge theory on manifolds with boundaries. Here we further the investigation undertaken in \cite{FP2018} by using the formalism of homotopy pullback and Deligne-Beilinson cohomology to describe an electromagnetic (EM) duality on the boundary of $M=B^{3}\times\mathbb{R}$. Upon breaking a generalized global symmetry, the duality is implemented by a BF-like topological boundary term. We then introduce Wilson line singularities on $\partial M$ and show that these induce the existence of dual edge modes, which we identify as connections over a $\left(-1\right)$-gerbe. We derive the pre-symplectic structure that yields the central charge in \cite{FP2018} and show that the central charge is related to a non-trivial class of the $\left(-1\right)$-gerbe.
}

\maketitle

%\tableofcontents

\section{Introduction}
\label{Section_1}

In recent years, there has been a lot of interest in the use of edge modes to construct an `extended phase space', which yields gauge-invariant symplectic structures and charges for gauge theories in the presence of a finite boundary. The central idea at the heart of this application of edge modes is that the boundary symmetry should not be conceived of as stemming from the `breaking of gauge symmetry' at the boundary, but rather as a true physical symmetry that is related to gauge-invariant observables.

The above series of investigations was initiated by Donnelly and Freidel in \cite{DF2016}, who used edge modes to define a boundary (from the perspective of a Cauchy surface) symplectic potential that compensates for the failure of gauge invariance of the bulk symplectic potential under field-dependent gauge transformations. More recently, Freidel and Pranzetti \cite{FP2018} attempted to extend the above analysis to construct EM duality \textit{only} on the boundary of a bulk spacetime, i.e. in the physically realistic case in which the bulk itself does not contain magnetic charges. Although such an EM duality was expected to hold from investigations into soft theorems and large gauge transformations in the asymptotic limit \cite{Strominger2016, Seraj2018}, Freidel and Pranzetti were the first to argue that one can construct this EM duality \textit{without} demanding that the large gauge transformations play the role of a physical symmetry generated by the charges. Furthermore, their analysis yields an important albeit heuristic result: The resulting magnetic and electric boundary charges do not commute, but instead give rise to a central charge. 

The reason that we call this result `heuristic' is that the analysis in \cite{FP2018} does not properly take into account the global structure of fields on the boundary, as well as the role that `singularities' play in the derivation of the central charge. 
One goal of this paper is thus to better explain the physical phenomenon of boundary EM duality by adopting a novel and systematic approach to the construction of edge modes that properly includes this global structure. 
However, our approach goes much further than merely justifying this result; it also allows us to answer some fundamental questions concerning the extended phase space construction from edge modes. For instance, what is the relationship between the original extended pre-symplectic structure derived in \cite{DF2016} and the distinct pre-symplectic structure that is necessary for the results of \cite{FP2018} to hold? Our framework provides an answer to this question by introducing a mechanism for spontaneous symmetry breaking on the boundary. Furthermore, we find interesting links between our construction and other topics that have seen a lot of recent interest, such as generalized global symmetries \cite{PS2005,PS2006a,PS2006b,Banks_2011,Gaiotto_2015}, duality walls \cite{Kapustin_2009}, the dressing field method \cite{Attard_2018}, and the appearance of higher structures (i.e. gerbes) \cite{fuchs2009bundle, Kapustin_2009} on the boundary of a field theory.  

In Section \ref{Section_2}, we review the construction of Freidel and Pranzetti \cite{FP2018} and note the necessity for a systematic explanation of the edge modes, as well as the construction of a pre-symplectic structure that reflects the non-commutation of boundary electric and magnetic charges (which is induced by singularities in the boundary sphere). 
In Section \ref{Section_3}, we lay the groundwork for a systematic construction of the relevant edge modes. In particular, we review our previous work \cite{Mathieu_2020}, where we showed that the edge modes in \cite{DF2016} can be constructed by means of two ingredients: At the kinematical level, they arise when one implements topological boundary conditions by means of a homotopy pullback, and at the dynamical level, the relevant bulk-to-boundary matching conditions and boundary symplectic structure arise from a particular choice of boundary action (on the latter, see also \cite{geiller2019extended}).\footnote{\cite{Mathieu_2020} also constructs the ghost and antifield phase space for the field content, but this will not be needed in the present work.} 

The subsequent sections then apply these tools to construct EM duality on the boundary and non-commuting electric and magnetic charges in codimension 2. 
Section \ref{Section_4} recalls that edge modes are intimately connected with the `boundary' spontaneous symmetry breaking of global 1-form symmetries, and Section \ref{Section_5} explains that the low-energy limit of the symmetry-breaking action yields a boundary action for EM duality. 
In order to motivate the introduction of the EM duality action that involves not only edge modes, but also \textit{dual} edge modes, Section \ref{Section_6} provides a description of regularized Wilson lines. Section \ref{Section_7} then introduces an edge mode version of the EM duality action that includes a dressed Wilson line on the boundary of the theory; it is the latter (codimension 2) structure that induces the dual edge modes. We give an explicit calculation showing how this results in the charge non-commutation relation produced in \cite{FP2018}, and note that the relevant boundary data can be formalized as a `non-trivial $\left(-1\right)$-gerbe with connection'. Finally, Section \ref{Section_8} generalizes this pattern to arbitrary dimensions and uses it to shed light on the (finite boundary) edge mode version of the asymptotic scalar-$2$-form duality discussed in \cite{Campiglia_2019}.

%Organization of paper:
%\begin{enumerate}

%\item Discuss problems with Freidel Pranzetti construction.

%\item Introduce MSTW edge modes and construction (without derived critical locus).

%\item Introduce our construction for FP case, and then for the case that goes down to points.

%\item Application: scalar-2-form duality
%\end{enumerate}

\section{Review of the Freidel-Pranzetti construction}
\label{Section_2}

In \cite{FP2018}, Freidel and Pranzetti (FP) work with the Lorentzian geometry $M=B^3 \times \mathbb{R}$ whose boundary is $\partial M = S^2 \times \mathbb{R}$. They then impose the action
\begin{equation}
\label{Action_FP}
    S = 
    \frac{1}{2e^{2}}\int_M F\wedge\star F 
    + \int_M A \wedge \star J 
    - \frac{1}{2\pi}\int_{\partial M} A\wedge d\tilde{a},
\end{equation}
where $J$ is assumed to be a gauge-invariant current, and $\tilde{a} := d\tilde{\phi} - \tilde{A}$, where $\tilde{\phi}$ is a scalar edge mode living on $\partial M$ which transforms as $\tilde{\phi} \mapsto \tilde{\phi} + \tilde{\epsilon}$ when $\tilde{A}$ undergoes the usual gauge transformation $\tilde{A} \mapsto \tilde{A} + d\tilde{\epsilon}$. $\tilde{\phi}$ is also known as the dressing field and the gauge-invariant quantity $\tilde{a}$ is the corresponding dressed field.

FP then introduce \textit{by hand} another edge mode $\phi \in \Omega^0 (\partial M)$ that is related to the gauge field $A$ and use $a:= d\phi - A$ to write down the `dressed form' of the pre-symplectic form that follows from the action \eqref{Action_FP}:
\begin{equation}
    \Omega 
    = \frac{1}{e^{2}}\int_\Sigma \delta A \wedge \star \delta F 
    + \frac{1}{2\pi}\int_{\partial \Sigma} \delta a \wedge \delta \tilde{a},
\end{equation}
where $\Sigma=B^3$ is a Cauchy surface at some point in time.

Finally, FP introduce what they call a `physical boundary symmetry' $\phi \mapsto \phi - \alpha, \tilde{\phi} \mapsto \tilde{\phi} - \tilde{\alpha}$ (an on-shell symmetry) which acts only on the edge modes, and use it to derive the (on-shell) charges
\begin{align}
    Q_E = \frac{1}{e^{2}}\int_{\partial \Sigma} \alpha *F 
    \quad\mbox{ and }\quad
    Q_M = 0, 
\end{align}
where $\Sigma$ is a spacelike Cauchy surface whose boundary $\partial\Sigma$ has the topology of $S^2$. Evidently, the commutator of these charges vanishes, but FP claim that upon imposing a `singularity' in the $A$ field on the boundary $S^2$ the on-shell magnetic charge should be modified to
\begin{equation}
    Q_M = \int_{S^1} \tilde{\alpha} a,
\end{equation}
where $S^1$ is a circle around the singular point on $S^2$, upon which the charge algebra is now centrally extended by 
\begin{equation}\label{central}
    \{ Q_E. Q_M \} = -\frac{1}{2\pi}\int_{S^1} \alpha d\tilde{\alpha}.
\end{equation}

While suggestive, this argument raises several questions. First, when and how can one introduce the edge modes in a fundamental way, instead of simply dressing `by hand' at the level of the pre-symplectic form? Second and more importantly, the central charge (\ref{central}) should arise from a particular pre-symplectic structure: What is this structure and how can one derive it from an action?
In what follows, we will try to answer these questions. We will see that in order to do so, one must be careful to construct the right gauge-invariant action from a generalized boundary condition, and be explicit about what one means by `singularities' in the boundary sphere. We now turn to the question of how the imposition of such boundary conditions is related to the existence of edge modes on the boundary. 

\section{The origin of edge modes: Homotopical boundary conditions}
\label{Section_3}

In \cite{Mathieu_2020}, we introduced the following way of imposing `relaxed' boundary conditions on the field content of a gauge theory. Consider a $\mathrm{U}\!\left(1\right)$ gauge theory on $M$ (which is contractible). The space of kinematic bulk fields can be formalized as $\mathrm{BU}\!\left(1\right)_{\text{con}}(M)$, the groupoid of principal $\mathrm{U}\!\left(1\right)$ bundles with connection on $M$\footnote{Without loss of generality, we should actually consider only \v{C}ech descriptions of bundles, as a groupoid has to be a small category.}. More concretely, since $M$ is contractible, we can think of this as the groupoid whose objects are globally defined gauge fields $A$ and whose morphisms are gauge transformations $A \mapsto A + d\chi$. We want to introduce a way of saying that the bulk fields restrict to a particular principal bundle (or principal bundle with connection) on $\partial M$; however, we must be careful here, because it would be too strict to require that the restricted bulk field \textit{equals} to some particular boundary field configuration; instead, the correct notion of comparison for gauge fields and bundles is not equality but \textit{isomorphism} --- as we are about to see, this simple point leads directly to the construction of edge modes on the boundary. 

In order to implement this relaxed notion of boundary condition, we use the following `homotopy pullback' diagram:
\begin{flalign}
\label{eqn:Boundarycondition1}
\xymatrix{
\ar@{-->}[d]\mathfrak{F}(M) \ar@{-->}[r] &  \ar@{}[dl]_-{h~~~~~} \mathrm{BU}\!\left(1\right)_{\text{con}}(M)\ar[d]^-{\mathrm{res}}\\
\{\ast\} \ar[r]_-{p}& \mathrm{BU}\!\left(1\right)\left(\partial M\right)
}
\end{flalign}
where $p$ is a functor picking out a particular principal bundle in $\mathrm{BU}\!\left(1\right)\left(\partial M\right)$. We note that in this particular diagram, we will impose the boundary condition of being a \textit{trivial bundle}; however, in general one can construct a diagram that includes non-trivial bundles and connections as boundary data if one likes, and we will find it necessary to include connection data in Section \ref{Section_6}. The field content resulting from imposing this boundary condition is derived by completing the pullback square to obtain the `homotopy pullback' $\mathfrak{F}(M)$.

In Appendix A of \cite{Mathieu_2020}, we provide a toolkit to compute the homotopy pullback $\mathfrak{F}\left(M\right)$ for general field content. In this section, on the other hand, our gauge fields are global objects so it suffices to adopt an elementary and hands-on approach to constructing the groupoid $\mathfrak{F}\left(M\right)$. (In Section \ref{Section_7}, we will have to be more careful about computing $\mathfrak{F}\left(M\right)$ because the boundary data is a non-trivial gerbe with connection; to perform this computation in a way that is accessible to physicists, we have chosen to use the Deligne-Beilinson presentation of differential cohomology as detailed in \cite{Bauer_2004}, see also Appendices \ref{Appendix_C} and \ref{Appendix_D} of the present article.) 

We now proceed to our elementary description of $\mathfrak{F}(M)$ for \eqref{eqn:Boundarycondition1}. First, an object of this groupoid is a pair $(A, \phi)$, where $\phi \in \Omega^{0}\left(\partial M\right)$ is a morphism that relates $p\left(\ast\right)$ to $\mathrm{res}\left(A\right)$, which is the boundary restriction of the bulk field. In other words, the `edge mode' $\phi$ witnesses the statement that the restriction of the bulk bundle is `the same as' (but not equal to) the trivial boundary bundle. 
Second, a morphism of this groupoid is a map that satisfies the following commuting diagram:
\begin{flalign}
\xymatrix@C=2.5em{
p\left(\ast\right) = \ast \ar[d]_-{\varphi} \ar[r]^-{\mathrm{id}_\ast}& \ast = p(\ast) \ar[d]^-{\varphi^\prime}\\
\mathrm{res}\left(A\right)=\ast \ar[r]_-{\varepsilon} & \ast = \mathrm{res}\left(A^\prime\right)
}
\end{flalign}
\
Thus, a morphism in $\mathfrak{F}\left(M\right)$ is given by $\left(A,\varphi\right) \stackrel{\epsilon}{\longrightarrow}\left(A+ d\epsilon , \varphi + \varepsilon\right)$, where $\epsilon \in \Omega^{0}\left(M\right)$. This is exactly the transformation law that \cite{DF2016} posit for edge modes when acted upon by the gauge symmetry. On the other hand, notice that the trivial boundary bundle $p\left(\ast\right)$ itself carries automorphisms that we can think of as an `external' transformation of the boundary condition: Under such an automorphism, $A$ is left invariant but $\phi\mapsto\phi - \alpha$, where $\alpha\in\Omega^{0}\left(\partial M\right)$. The latter is precisely the transformation law for $\phi$ under what \cite{DF2016, FP2018} call the \textit{physical} boundary symmmetry (as opposed to the gauge symmetry) of the edge mode. In other words, our construction shows that edge modes come from a particular \textit{boundary condition} and that once one understands that boundary structure, one obtains the so-called `physical boundary symmetries' for free.
Finally, we emphasize that on the boundary $\partial M$, the edge modes can be used to `dress' the gauge field, yielding the definition of the dressed photon $a:= d\phi - A$.

\section{A Higgs model from edge modes}
\label{Section_4}

We recall that \cite{DF2016} introduces a boundary pre-symplectic form $\Omega_{\partial\Sigma}$ that compensates for the failure of gauge invariance of the standard pre-symplectic form $\Omega_{\Sigma}$ under field-dependent gauge transformations in the presence of a finite boundary. For instance, in the case of electromagnetism, the standard pre-symplectic form is $\Omega_{\Sigma} = \int_{\Sigma} \star\delta F \wedge \delta A$ and the boundary pre-symplectic form given in \cite{DF2016} is $\int_{\partial \Sigma} \star \delta F \wedge \delta \phi$. Essentially, their construction proceeds by dressing the standard presymplectic form (which ensures gauge invariance), and noticing that this leads to a boundary contribution. They also use gauge invariance to motivate the bulk-to-boundary `matching condition' $\left.\star F\right|_{\partial M} = \star_{\partial} a$, where $\star_{\partial}$ denotes the Hodge star operator with respect to the boundary $\partial M$.

In the previous section, we showed that at the kinematic level, such a dressing field $\phi$ naturally arises on the boundary by implementing our homotopical boundary conditions. However, according to the covariant phase space formalism, we need \textit{dynamical input} in order to define the boundary pre-symplectic form corresponding to such a $\phi$: It should arise from a well-defined variational problem on the boundary. In \cite{Mathieu_2020} (see also \cite{geiller2019extended}) we derive the boundary pre-symplectic form given in \cite{DF2016} as well as their bulk-to-boundary matching condition (indeed, we derive the symplectic structure of the entire BRST extension of the theory) from the following action:

\begin{equation}\label{MMSPaction}
    S 
    = \frac{1}{2e^{2}}\int_M F \wedge \star F 
    + \frac{t^{2}}{2}\int_{\partial M} a \wedge \star_{\partial} a,
\end{equation}
where $\star_\partial$ is the Hodge star on $\partial M$ and $a:= d\phi - A$ can either be interpreted as the dressing of $A$, or as an affine covariant derivative $d_A \phi$. The equations of motion resulting from varying this action are 
\begin{align}
\star F =& 0 \mbox{ on } M,\\
d \star a =& 0 \mbox{ on } \partial M,\\
\left.\star F\right|_{\partial M} =& \star_{\partial} a \mbox{ on } \partial M;
\end{align}
and the presymplectic $2$-form is 
\begin{equation}
\Omega = \int_{\Sigma} \star \delta F \wedge \delta A + \int_{\partial \Sigma} \star_{\partial} \delta a \wedge \delta \phi.
\end{equation}
We note that these structures are of course gauge-invariant. 

When we view $a$ as $d_{A}\phi$, it is clear that $t^{2} a \wedge \star_{\partial} a$ provides a description of the Higgs phase of the theory, i.e. it is the kinetic term for a charged scalar coupled to our $\mathrm{U}\!\left(1\right)$ gauge theory. In other words, on the \textit{boundary} $\partial M$, the $\mathrm{U}\!\left(1\right)$ gauge symmetry of the bulk Maxwell action has been Higgsed to a $\mathbb{Z}$ gauge symmetry. On the other hand, if we expand the boundary action in terms of $A$ and $\phi$, we can also think of it as the Proca action for a massive vector field $A$, where a Stuckelberg field $\phi$ has been introduced to maintain the gauge-invariance.\footnote{For an attempt to explain why such Stuckelberging should generically occur at the boundary of a subsystem, we refer the reader to \cite{Dvali_2016}.}

It is instructive to ask what happens to the \textit{global} $1$-form symmetries of a pure Maxwell gauge theory on $M$ when we couple it to the boundary action. We recall \cite{Gaiotto_2015} that in a free Maxwell theory, there are two global $1$-form symmetries, viz. the electric and magnetic ones. The electric charge operator is $\int_{\mathcal{C}} \star F$ and the magnetic charge operator is $\int_{\mathcal{C}} F$, where $\mathcal{C}$ is a surface.
%\textcolor{red}{What is the degree of $j_{E}$ and $j_{M}$? Since $\int_{\mathcal{C}} F$ and $\int_{\mathcal{C}}\star F$ are numbers (provided $M$ is $4$-dimensional), then $j_{E}$ and $j_{M}$ have to be $4$-forms so that $\star j$ is a $0$-form, but still, we want a number, not a function, so I think it is incorrect. I think what you mean is, by the eom, $\int_{V}d\star F = \int_{V}\star j_{E} = Q^{E}$ which leads by Stokes to $\int_{\partial V = S}\star F = Q^{E}$ and you want, by analogy, to write $\int_{\partial V = S}\tilde{F} = Q^{M}$ which is weird, as if you use Stokes backwards, you get $\int_{V}d\tilde{F} = Q^{M}$ ie $Q^{M} = 0$ since $dF = 0$ by definition of the curvature of a connection. I've looked at reference [5] and they do not address the issue. In particular, they are rather sloppy on the definition of their surface $M^{2}$ (or maybe I've missed something). I guess what they mean is that the surface should be topologically nontrivial in order to have a nonzero symmetry. Anyway, there is something wrong in the draft here. Moreover, there is probably a normalization to check here but maybe we don't really care.} 
As discussed in \cite{Gaiotto_2015}, if one then goes on to Higgs the pure Maxwell theory by coupling it to a charge $N$ scalar, we break the global $1$-form \textit{electric} symmetry down to a $\mathbb{Z}_{N}$ symmetry (so our case is simply the $N=1$ case). The difference between the scenario discussed in \cite{Gaiotto_2015} and ours is that for us this Higgsing only happens on the boundary $\partial M$, where we do not need to postulate the existence of the Higgs field $\phi$, because it arises for free from the homotopical boundary condition \eqref{eqn:Boundarycondition1}. One expects \cite{lake2018higherform} that the breaking of a $1$-form global symmetry gives rise to a $1$-form Goldstone mode, and indeed the boundary term $\int_{\partial M} a \wedge\star_{\partial}a$ is identified in \cite{Hofman_2019} as the Goldstone action. In other words, the emergent gauge-invariant degree of freedom is simply the dressed photon $a$, which transforms under the physical boundary symmetry as $a - d\alpha$. 

\section{EM duality in the low-energy limit}
\label{Section_5}

In the previous section, we established that Higgsing a $\mathrm{U}\!\left(1\right)$ theory \textit{on the boundary} leads to the extended phase space (and bulk-to-boundary matching conditions) invoked in the seminal paper \cite{DF2016} connecting edge modes to the symplectic structure of gauge theories.
However, the EM duality version of this construction \cite{FP2018} uses a different action, i.e. action \eqref{Action_FP}, from the action given in \cite{DF2016}, i.e. action \eqref{MMSPaction}.
What then is the connection between these two models (i.e. action \eqref{MMSPaction} and action \eqref{Action_FP})? 
We note that the key property of \eqref{Action_FP} is that it has a BF boundary term \footnote{By ``BF term'' we mean here a term whose Lagrangian is $B\wedge F = B\wedge dA$ where $A$ \textbf{and} $B$ are $\mathrm{U}\!\left(1\right)$ connections, contrary to the more common case where $B$ is simply a global background differential form.}, and thus the connection is given by a standard argument \cite{Banks_2011} showing that a BF term appears when we take the low energy limit ($t^2 \rightarrow \infty$) of the boundary term in \eqref{MMSPaction}. For completeness, we briefly review this argument in our context. 

Consider again the boundary term $\int_{\partial M} a \wedge \star_{\partial} a$ in \eqref{MMSPaction}. We now dualize the edge mode $\phi$ by introducing a $2$-form Lagrange multiplier $H$ and writing the boundary action as
\begin{equation}
    S_{\partial} = \frac{1}{2t^2}\int_{\partial M} H \wedge \star_{\partial} H + \int_{\partial M} H \wedge a,
\end{equation}
where the equivalence is demonstrated by varying $H$ to yield the equation of motion 
\begin{equation}
\star_{\partial}H = t^{2}a \mbox{ on } \partial M.    
\end{equation}
If we then integrate our edge mode $\phi$ from this form of the boundary action and take the low-energy limit $t^{2}\rightarrow\infty$, we obtain the low-energy boundary action $\tilde{A} \wedge F$, where $\tilde{A}$ is a $1$-form such that $H = d\tilde{A}$.

%\textcolor{red}{Be careful that BF theory can mean several things. Usually, the $B$ field is a global background form that does not transform. Here, it is a connection.} \textcolor{blue}{For convenience we assume that $\tilde{A}$ is a global $1$-form so it corresponds to a trivial connection.} \textcolor{red}{It does not solve the issue. In the usual SU(2) theory, we have $A\rightarrow g^{-1}Ag + g^{-1}dg$ so $F\rightarrow g^{-1}Fg$ and $B\rightarrow g^{-1}Bg$. Therefore in the Abelian case we have $A\rightarrow A + d\lambda$ so $F\rightarrow F$ and $B\rightarrow B$. So you see that the nature of $A$ and $B$ are truly different, even in the Abelian case, and even if we assume $A$ is a trivial connection.}

This low-energy boundary action looks like a BF action in 3D, but as we are about to see, it differs from a free BF theory whose equations of motion yield the flatness of $A$ and $\tilde{A}$. The new total (bulk plus boundary) action is 
\begin{equation}\label{BF1}
    S = \frac{1}{2e^{2}}\int_{M} F \wedge \star F + \frac{k}{2\pi}\int_{\partial M}\tilde{A} \wedge F,
\end{equation}
which yields the equation of motion 
\begin{equation}
\left.\star F\right|_{\partial M} = \tilde{F}:= d\tilde{A} \mbox{ on } \partial M.    
\end{equation}
upon varying $A$, as well as a gauge-invariant presymplectic potential and presymplectic form. 
We stress two points about this model. First, the equation of motion that matches bulk to boundary fields is precisely the source of the boundary EM duality. It is amusing to note that this is nothing other than an elementary implementation of the duality wall formalism discussed in \cite{Kapustin_2009}: There, the prescription for witnessing a duality is to introduce a `wall operator' on codimension $1$ surface separating the dual theories, and in the case of the S-transformation of EM duality, that wall operator takes the form of a BF term. Second, the gauge-invariance of the presymplectic structure relies on the fact that $\partial M$ is closed and that there is no higher codimension stratum that would spoil the gauge-invariance via Stokes' theorem (thus obviating the need to introduce edge modes). However, as we will soon see, in order to recover a non-commutation relation between electric and magnetic charges along the lines of \cite{FP2018}, we need to introduce not only the edge modes coming from placing a (homotopical) boundary condition on the bulk $\mathrm{U}\!\left(1\right)$ bundle with connection $A$, but also the \textit{dual} edge modes $\tilde{\phi}$ coming from placing a boundary condition on the \textit{boundary} dual bundle with connection $\tilde{A}$. In other words, we expect the dual edge modes to arise on a codimension $2$ stratum (with respect to $M$) that lies in the boundary $\partial M$. In the next section, we discuss how such a stratum emerges from inserting electric Wilson lines into the boundary, and in Section \ref{Section_7} we use these edge modes to write down a dressed action that leads to the desired non-commutation relation. We will see that this dressed action is equivalent to supplementing the action (\ref{BF1}) with some codimension two data.

%Again it is instructive here to inquire how our low-energy action $(\ref{BF1})$ has broken the global $1$-form symmetries of a free $U(1)$ BF theory in 3 dimensions. We recall that a free 3D BF theory has electric (or dual magnetic) and magnetic $\mathbb{Z}$ $1$-form symmetries, whose charge operators are $\int \tilde{A} = \star j_E$ and $\int A$ respectively. We thus see that the equation of motion $\star F = \tilde{F}$ breaks the electric $1$-form symmetry. There should be a $1$-form Goldstone mode associated with this breaking, but since $\partial M$ is closed, we find that if we dress $\tilde{A}$ by an edge mode $\tilde{\phi}$ to form $\tilde{a} = d\tilde{\phi} - \tilde{A}$ the resulting term vanishes by Stokes' theorem. Evidently the $\tilde{\phi}$ edge mode will only come into its own when there is a codimension 2 stratum in the theory. As we will now see, such codimension 2 structure is precisely what we need in order to formulate edge mode duality and witness the presence of a central charge in the theory.

\section{Wilson lines for the dressed gauge field}
\label{Section_6}

We recall that according to \cite{FP2018}, the source of the non-trivial boundary magnetic charge and the central charge can be traced to the existence of singularities piercing the boundary sphere $S^2$. From the perspective of our boundary BF theory, such singularities can be thought of as probe electric Wilson lines for the dressed photon $a$ that are extended in time, i.e. as $\int_{\gamma} a$ along a timelike curve $\gamma = \left\lbrace p_{t}\in S^{2}\times\left\lbrace t\right\rbrace,\,t\in\mathbb{R}\right\rbrace\subset\partial M$. The curve $\gamma$ can be visualized as piercing the spatial boundary $S^{2}\times\left\lbrace t\right\rbrace$ at $p_{t}$ for any time $t\in\mathbb{R}$.

Following \cite{Kapustin_2009}, we now introduce a regularized description of such a Wilson line into our theory. Let $Z^{\varepsilon}_{\gamma} = \left\lbrace B^{2}_{\varepsilon}\left(p_{t}\right),\,t\in\mathbb{R}\right\rbrace$ be a tubular neighborhood of $\gamma$ where $B^{2}_{\varepsilon}\left(p_{t}\right)$ is the $2$-ball of radius $\varepsilon$ centered at $p_{t}\in S^{2}\times\left\lbrace t\right\rbrace$. Hence, $\partial Z^{\varepsilon}_{\gamma} = \left\lbrace S^{1}_{\varepsilon}\left(p_{t}\right),\,t\in\mathbb{R}\right\rbrace$. 

We can then construct a closed and co-closed $1$-form $\Omega_{n}$ on $\Delta^{\varepsilon}_{\gamma} = \partial M\setminus \overset{\circ}{Z^{\varepsilon}_{\gamma}}$ (thus $\Delta^{\varepsilon}_{\gamma} \cong B^{2}\times\mathbb{R}$), with integral periods (up to a $2\pi$ factor) such that, for a fixed integer $n$:
\begin{enumerate}[i)]
    \item $\left.\star_{\partial}\Omega_{n}\right|_{\partial Z^{\varepsilon}_{\gamma}} = 0$,
    \item $\int_{S^{1}_{\varepsilon}\left(p_{t}\right)} \Omega_{n} = 2\pi n$ for any $t\in\mathbb{R}$.
\end{enumerate}
The existence of such a $1$-form $\Omega_{n}$ is proven in \cite{Kapustin_2009}. A key point of the proof is the fact that the linking number between $\gamma$ and $S^{1}_{\varepsilon}\left(p_{t}\right)$ is $1$ for any $t\in\mathbb{R}$ by definition.

The regularized Wilson line can then be defined as 
\begin{equation}\label{DressedWilson}
    W_{\varepsilon}(\gamma) = \int_{\Delta^{\varepsilon}_{\gamma}} a \wedge \Omega_{n}.
\end{equation}
This picture can be generalized to the case of multiple Wilson lines. In what follows, we will drop the index $n$ of $\Omega_{n}$ and the superscripts and subscripts of $\Delta_{\gamma}^{\varepsilon}$.

\section{Dual edge modes and the central charge}
\label{Section_7}

In this section, we want to start with our homotopy pullback construction and produce the kinematic resources to write down an action that 
\begin{enumerate}
\item Accommodates dressed boundary Wilson lines that are analogous to (\ref{DressedWilson}).
\item Reproduces the EM duality on $\partial M$.
\item Gives rise to \textit{non-commuting} electric and magnetic charges defined in terms of edge modes as conjectured in \cite{FP2018}.
\end{enumerate}
We will do so by initially working with dressed fields on the boundary, since this guarantees the gauge invariance of the action under non-trivial boundary gauge transformations; but as we will soon see, it also leads to an interesting reformulation in terms of undressed variables. 

From Section \ref{Section_6}, we know that in order to include regularized Wilson lines of the form $\int_\Delta A \wedge \Omega$, we need to consider a submanifold $\Delta \subset \partial M$ whose boundary $\partial \Delta$ is homeomorphic to $S^1 \times \mathbb{R}$. We thus proceed to formulate our boundary bundles and action terms on $\Delta$ and $\partial\Delta$. It will be convenient for us to abuse language by also using $\Delta$ to refer to the smallest open set (with respect to inclusion) containing $\Delta$ (and thus $\partial\Delta$) as it is defined above. 

To be clear, there are two kinds of edge modes that we wish to construct in this section. First, the standard edge mode $\phi$ on the codimension $1$ submanifold $\partial M$ is constructed in the same way as \eqref{eqn:Boundarycondition1}, i.e. by means of the following homotopy pullback square:
\begin{flalign}\label{eqn:Boundarycondition2}
\xymatrix{
\ar@{-->}[d]\mathfrak{F}(M) \ar@{-->}[r] &  \ar@{}[dl]_-{h~~~~~} \mathrm{BU}\!\left(1\right)_{\text{con}}(M)\ar[d]^-{\mathrm{res}}\\
\{\ast\} \ar[r]_-{p}& \mathrm{BU}\!\left(1\right)\left(\partial M \right)
}
\end{flalign}

Second, as we argued at the end of Section \ref{Section_5}, the \textit{dual} edge modes $\tilde{\phi}$ should come from placing a codimension $2$ boundary condition on the boundary dual bundle that lives on $\Delta$. Thus, in this scenario the existence of the dual edge modes is actually induced by the regularized Wilson line, and we will take the relevant codimension $2$ surface to be $\partial \Delta$. More precisely, we will understand $\tilde{\phi}$ as arising from the following homotopy pullback:
\begin{flalign}\label{eqn:Gboundarycondition}
\xymatrix{
\ar@{-->}[d]\tilde{\mathfrak{F}}\left(\Delta\right) \ar@{-->}[r] &  \ar@{}[dl]_-{h~~~~~} \mathrm{BU}\!\left(1\right)_{\text{con}}(\Delta)\ar[d]^-{\mathrm{res}}\\
\{\ast\} \ar[r]_-{p}& \mathrm{BU}\!\left(1\right)^{-}_{\text{con}}(\partial \Delta)
}
\end{flalign}
where we stress that we are now imposing a new kind of boundary condition by means of the groupoid $\mathrm{BU}(1)^{-}_{\text{con}}$. We will describe this groupoid explicitly in just a moment, but the important thing to note is that it restricts the field content to a particular principal bundle on $\partial \Delta$ (thus also restricting the transformations which preserve that data) and in addition incorporates a connection on that bundle (which is natural because the Wilson line data that induces $\partial\Delta$ involves a connection).

In order to compute $\tilde{\mathfrak{F}}\left(\Delta\right)$, we will first need to properly define the elements at the other corners of diagram \eqref{eqn:Gboundarycondition}. In what follows, we will assume that $\Delta$ is provided with a good cover $\mathcal{U}=\left\lbrace U_{i}\right\rbrace_{i}$ (i.e. the open sets of $\mathcal{U}$ and their intersections are contractible or empty) which induces a good cover $\mathcal{V}=\left\lbrace V_{i}\right\rbrace_{i}$ on $\partial\Delta$.

First, $\mathrm{BU}\!\left(1\right)_{\mathrm{con}}\left(\Delta\right)$ is the groupoid of $\mathrm{U}\!\left(1\right)$ connections over $\Delta$. We remind the reader that a $\mathrm{U}\!\left(1\right)$-connection $\tilde{A}$ is a collection of local $1$-forms $\tilde{A}_{i}$ in the $U_{i}$'s, a collection of $\tilde{\Lambda}_{ij}$ in the double intersections $U_{ij}=U_{i}\cap U_{j}$ (i.e. the arguments of the transition functions of the $\mathrm{U}\!\left(1\right)$-bundle over which the connection is defined) and a collection of elements $\tilde{n}_{ijk}\in 2\pi\mathbb{Z}$ in the triple intersections $U_{ijk}=U_{i}\cap U_{j}\cap U_{k}$ (which define a \v{C}ech $2$-cocyle indicating the isomorphism class of the $\mathrm{U}\!\left(1\right)$-bundle over which the connection is defined) satisfying the following so-called descent equations:
\begin{align}
\begin{cases}
    &\tilde{A}_{i}-\tilde{A}_{j} - d\tilde{\Lambda}_{ij} = 0 \mbox{ in }\,U_{ij}\\
    &\tilde{\Lambda}_{ij}-\tilde{\Lambda}_{ik}+\tilde{\Lambda}_{jk} - \tilde{n}_{ijk}=0 \mbox{ in }\,U_{ijk}\\
    &\tilde{n}_{ijk}-\tilde{n}_{ijl}+\tilde{n}_{ikl}-\tilde{n}_{jkl} = 0
\end{cases}
\end{align}

More concisely, we can write this as
\begin{equation}
    D^{\left[1,1\right]}\tilde{A} = 0
\end{equation}
by means of the Deligne-Beilinson differential
\begin{equation}
\label{D_cocycle_1}
    D^{\left[1,1\right]} =
    \begin{pmatrix}
    \check{\delta} & -d & 0 \\
    0 & \check{\delta} & -d \\
    0 & 0 & \check{\delta} \\
    \end{pmatrix}
\end{equation}
where $\check{\delta}$ is the \v{C}ech coboundary operator and $d$ is the usual de Rham differential operator (which simply injects integers into the set of constant $\mathbb{Z}$-valued functions over $U_{ijk}$ when applied to $\left(\tilde{n}_{ijk}\right)$) and
\begin{equation}
    \tilde{A} =
    \begin{pmatrix}
    \left(\tilde{A}_{i}\right)_{i} \\
    \left(\tilde{\Lambda}_{ij}\right)_{ij} \\
    \left(\tilde{n}_{ijk}\right)_{ijk}
    \end{pmatrix}.
\end{equation}
We say that $\tilde{A}$ is a Deligne-Beilinson (DB) cocycle of degree $1$, referring to the form degree of the local fields $\tilde{A}_{i}$, or a differential cocycle of degree $2$ (where the latter convention makes it possible to have a cup product that is truly graded commutative). The $\mathbb{Z}$-module of DB $1$-cocycles over $\Delta$ will be denoted as $Z^{1}_{\mathrm{DB}}\left(\Delta\right)$.

A general `gauge transformation' of $\tilde{A}$ is given by the following set of relations:
\begin{align}
\label{Descent_Eq_GT}
\begin{cases}
    &\tilde{A}_{i}\longrightarrow\tilde{A}_{i} + d\tilde{q}_{i} \mbox{ in }\,U_{i}\\
    &\tilde{\Lambda}_{ij}\longrightarrow\tilde{\Lambda}_{ij} + \tilde{q}_{i} - \tilde{q}_{j} + \tilde{m}_{ij} \mbox{ in }\,U_{ij}\\
    &\tilde{n}_{ijk}\longrightarrow\tilde{n}_{ijk} + \tilde{m}_{ij} - \tilde{m}_{ik} + \tilde{m}_{jk} \mbox{ in }\,U_{ijk}
\end{cases}
\end{align}
or more concisely,
\begin{equation}
    \tilde{A} \longrightarrow \tilde{A} + D^{\left[1,0\right]}\tilde{q}
\end{equation}
with 
\begin{equation}
\label{D_coboundary_0}
    D^{\left[1,0\right]} =
    \begin{pmatrix}
    d & 0 \\
    \check{\delta} & d \\
    0 & \check{\delta} \\
    \end{pmatrix}
\end{equation}
(note that $D^{\left[1,1\right]}\circ D^{\left[1,0\right]} = 0$) and
\begin{equation}
    \tilde{q} =
    \begin{pmatrix}
        \left(\tilde{q}_{i}\right)_{i} \\
        \left(\tilde{m}_{ij}\right)_{ij}
    \end{pmatrix}
\end{equation}
As one can see from equations \eqref{Descent_Eq_GT}, this type of gauge transformation corresponds to simultaneously performing a bundle isomorphism (thus changing the representative of a bundle isomorphism class) and a change of section. 
We say that $\tilde{q}$ is a DB $0$-cochain (the $\mathbb{Z}$-module of DB $0$-cochains over $\partial\Delta$ being denoted $C^{0}_{\mathrm{DB}}\left(\partial\Delta\right)$) and $D^{\left[1,0\right]}\tilde{q}$ is a DB coboundary of degree $1$ (the $\mathbb{Z}$-module of DB $1$-coboundaries over $\partial\Delta$ being denoted $B^{1}_{\mathrm{DB}}\left(\Delta\right)$).

In our present language, the groupoid of $\mathrm{U}\!\left(1\right)$-connections over $\Delta$ can now be described in terms of the following objects and morphisms:
\begin{align}
\nonumber
    \mathrm{BU}&\left(1\right)_{\mathrm{con}}\left(\Delta\right)=
    \begin{cases} 
        \mathrm{Obj:}\,\tilde{A}\in Z^{1}_{\mathrm{DB}}\left(\Delta\right) \\
        \mathrm{Mor:}\,\tilde{A} \overset{\tilde{q}}{\longrightarrow}\tilde{A} + D^{\left[1,0\right]}\tilde{q},\,\tilde{q}\in C^{0}_{\mathrm{DB}}\left(\Delta\right)
    \end{cases}
\end{align}
This specifies the data of the upper right corner in diagram \eqref{eqn:Gboundarycondition}.

Actually, we would also like to consider a more restricted scenario in order to specify the data of the lower right corner in \eqref{eqn:Gboundarycondition}. Notice that if
\begin{equation}
\label{Connection_-1_Gerbe}
    D^{\left[0,0\right]}\tilde{q} = 0
\end{equation}
where
\begin{equation}
\label{D_cocycle_0}
    D^{\left[0,0\right]} =
    \begin{pmatrix}
    \check{\delta} & d \\
    0 & \check{\delta} \\
    \end{pmatrix}
\end{equation}
then the gauge transformation $\tilde{A}\longrightarrow \tilde{A} + D^{\left[1,0\right]}\tilde{q}$ corresponds to a change of section on the same bundle. The difference between \eqref{D_coboundary_0} and \eqref{D_cocycle_0} is a matter of truncation of the \v{C}ech-de Rham complex for the construction of DB complex, see \cite{Bauer_2004} and Appendix \ref{Appendix_C} for more details. When \eqref{Connection_-1_Gerbe} is satisfied, we say that $\tilde{q}$ is a $0$-cocycle and use $Z^{0}_{\mathrm{DB}}\left(\Delta\right)$ to denote the $\mathbb{Z}$-module of $0$-cocycles over $\Delta$. Concretely, $\tilde{q}$ is simply a collection of smooth real-valued functions over the open sets of the (good) cover of $\Delta$ such that, over every non-empty double intersection, the functions over each of the two open sets agree up to an element of $2\pi\mathbb{Z}$. In other words, the complex exponential of the local functions is a globally well-defined $\mathrm{U}\!\left(1\right)$-valued map over $\Delta$. In more mathematical terms, such an object $\tilde{q}$ can be referred to as a \textit{connection over a} $\left(-1\right)$-\textit{gerbe} over $\Delta$, see Appendix \ref{Appendix_D} for more details. 

The data for the lower right corner in \eqref{eqn:Gboundarycondition} can now be concisely described as the groupoid $\mathrm{BU}\!\left(1\right)^{-}_{\mathrm{con}}\left(\partial\Delta\right)$:
\begin{align}
\nonumber
    \mathrm{BU}\!\left(1\right)^{-}_{\mathrm{con}}\left(\partial\Delta\right)=
    \begin{cases} 
        \mathrm{Obj:}\,\tilde{B}\in Z^{1}_{\mathrm{DB}}\left(\partial\Delta\right) \\
        \mathrm{Mor:}\,\tilde{B} \overset{\tilde{\phi}}{\longrightarrow}\tilde{B}+D^{\left[1,0\right]}\tilde{\phi},\,\tilde{\phi}\in Z^{0}_{\mathrm{DB}}\left(\partial\Delta\right)
    \end{cases}
\end{align}
where the morphisms are connections over $\left(-1\right)$-gerbes. Note that this groupoid is equivalent to the groupoid whose objects are $\mathrm{U}\!\left(1\right)$-valued functions and whose morphisms are just the identity.

We are finally ready to compute the homotopy pullback $\tilde{\mathfrak{F}}\left(\Delta\right)$ of the diagram \eqref{eqn:Gboundarycondition}. First recall that the functor $p$ just selects an element of $\mathrm{BU}\!\left(1\right)^{-}_{\mathrm{con}}\left(\partial\Delta\right)$, while the functor $\mathrm{res}$ restricts to $\partial\Delta$ the element of $\mathrm{BU}\!\left(1\right)_{\mathrm{con}}\left(\Delta\right)$, which is defined over $\Delta$. Then, by definition of the homotopy pullback, the objects of $\tilde{\mathfrak{F}}\left(\Delta\right)$ are triplets $\left(\ast,\tilde{A},\tilde{\phi}\right)\in\left\lbrace*\right\rbrace\times Z^{1}_{\mathrm{DB}}\left(\Delta\right)\times Z^{0}_{\mathrm{DB}}\left(\partial\Delta\right)$ (and we will omit the first component $\ast$ in the following). A morphism $\left(\tilde{A},\tilde{\phi}\right)\longrightarrow\left(\tilde{A}',\tilde{\phi}'\right)$ in $\tilde{\mathfrak{F}}\left(\Delta\right)$ is a pair of morphisms $\left(\mathrm{id}_{\ast}: \ast \longrightarrow \ast, \varepsilon: A \longrightarrow \tilde{A}' = \tilde{A} + D^{\left[1,0\right]}\tilde{\varepsilon}\right)$ that is compatible with $\tilde{\phi}$ and $\tilde{\phi}'$, i.e. such that the diagram
\begin{flalign}\label{eqn:Boundarycondition3}
\xymatrix{
\ar[d]_-{\tilde{\phi}} \ast \ar[r]^-{\mathrm{id_{\ast}}} &  \ar[d]^-{\tilde{\phi}'} \ast \\
\mathrm{res}\left(\tilde{A}\right) \ar[r]_-{\tilde{\varepsilon}}& \mathrm{res}\left(\tilde{A}'\right)
}
\end{flalign}
commutes. Hence, a morphism in $\tilde{\mathfrak{F}}\left(\Delta\right)$ is given by $\left(\tilde{A},\tilde{\phi}\right)\longrightarrow\left(\tilde{A}'= \tilde{A} + D^{\left[1,0\right]}\tilde{\varepsilon},\tilde{\phi}' = \tilde{\phi} + \tilde{\varepsilon}\right)$, where $\tilde{\varepsilon}\in Z^{0}_{\mathrm{DB}}\left(\Delta\right)$ (and not simply $C^{0}_{\mathrm{DB}}\left(\Delta\right)$ since both $\tilde{\phi}$ and $\tilde{\phi}'$ are $0$-cocycles) and we can rewrite
\begin{align}
\nonumber
    \tilde{\mathfrak{F}}\left(\Delta\right)=
    \begin{cases} 
        \mathrm{Obj:}\,\left(\tilde{A},\tilde{\phi}\right)\in Z^{1}_{\mathrm{DB}}\left(\Delta\right)\times Z^{0}_{\mathrm{DB}}\left(\partial\Delta\right) \\
        \mathrm{Mor:}\,\left(\tilde{A},\tilde{\phi}\right) \overset{\tilde{\varepsilon}}{\longrightarrow}\left(\tilde{A}+D^{\left[1,0\right]}\tilde{\varepsilon},\tilde{\phi}+\tilde{\varepsilon}\right),\,\tilde{\varepsilon}\in Z^{0}_{\mathrm{DB}}\left(\Delta\right)
    \end{cases}
\end{align}

Now, we can introduce the covariant derivative of $\tilde{\phi}$:
\begin{equation}
    D_{A}\phi = D^{\left[1,0\right]}\tilde{\phi} - \tilde{A}
\end{equation}
or, written differently:
\begin{align}
    D_{A}\tilde{\phi} 
    =
    \begin{pmatrix}
    d & 0 \\
    0 & 0 \\
    0 & 0 \\
    \end{pmatrix}
    \begin{pmatrix}
    \left(\tilde{\phi}_{i}\right)_{i} \\
    \left(\tilde{m}_{ij}\right)_{ij} \\
    \end{pmatrix}
    - \begin{pmatrix}
    \left(\tilde{A}_{i}\right)_{i} \\
    \left(\tilde{\Lambda}_{ij}\right)_{ij} \\
    \left(\tilde{n}_{ijk}\right)_{ijk} \\
    \end{pmatrix}
    \label{Covariant_Derivative}
    =
    \begin{pmatrix}
    \left(d\tilde{\phi}_{i} - \tilde{A}_{i}\right)_{i} \\
    \left(- \tilde{\Lambda}_{ij}\right)_{ij} \\
    \left(- \tilde{n}_{ijk}\right)_{ijk} \\
    \end{pmatrix}
\end{align}
since $D^{\left[0,0\right]}$ acts by definition like the null operator on $Z^{0}_{\mathrm{DB}}\left(\partial\Delta\right)$ and can be regarded as a submatrix of $D^{\left[1,0\right]}$.

At this point, several remarks are in order. First, recall that the $n$-gerbes over a manifold $Y$ are classified up to isomorphism by $H^{n+2}\left(Y;\mathbb{Z}\right)$. So here, the $\left(-1\right)$-gerbes over $\partial\Delta$ are classified up to isomorphism by $H^{1}\left(\partial\Delta;\mathbb{Z}\right) \cong H^{1}\left(S^{1};\mathbb{Z}\right)\cong\mathbb{Z}$. Hence, there are nontrivial $\left(-1\right)$-gerbes and as a consequence, there are some connections $\tilde{\phi}$ over such $\left(-1\right)$-gerbes that are not gauge equivalent to a $\mathbb{R}$-valued function (the exponent of the transition function) globally defined over $\partial\Delta$. In fact, once we introduce an appropriate action, the cohomology class in $H^{1}\left(\partial\Delta;\mathbb{Z}\right)$ that indexes the $\left(-1\right)$-gerbes is constrained on-shell by the equations of motion, as we will soon see. Second, in the following, we want to assume that $\tilde{A}$ is globally defined. Hence, the first component of $D_{A}\phi$ is globally defined also (recall that since $D^{\left[0,0\right]}\tilde{\phi} = 0$, the functions $\tilde{\phi}_{i}$ in the $U_{i}$'s differ by integers in the $U_{ij}$'s). From now on, we will denote this `dressed' quantity as $\tilde{a}$:
\begin{equation}
    \tilde{a} := d\tilde{\phi}_{i} - \tilde{A}
\end{equation}
This will be our main (gauge-invariant) building block in the following. 

%Since we would like to formulate EM duality witnessed by the 

%Earlier we saw that our edge modes $\phi$ arose from homotopically imposing a codimension 1 boundary condition on the bulk $U(1)$ bundle. Upon taking the low-energy limit of the bounndary action, we then found that we also had a dual bundle $\tilde{P}$ with connection $\tilde{A}$ defined on the codimension 1 stratum $\partial M$. In order to obtain dual edge modes $\tilde{\phi}$, we would thus expect to impose a homotopical boundary condition on some codimension 2 substratum. We now explain how to motivate such a boundary condition. 

%The submanifold $\partial \Delta$ will play the role of a codimension 2 stratum in the new theory that we are about to introduce, and so we would like to implement on it a boundary condition for the dual field $(\tilde{P}, \tilde{A})$ that lives on the codimension 1 stratum $\Delta \subset \partial M$. To do so, we draw the following homotopy pullback diagram: 
%Following the prescription of *** to compute yields. Morphisms. (We assume that the edge modes $\phi$ exist because of the homotopy pullback ***)

We have now completed the computation of the field content arising from the homotopy pullback, and are ready to use it in the following manifestly gauge-invariant action:
\begin{align}
    \label{Action_4D}
    S = \frac{1}{2e^{2}}\int_M F \wedge \star F 
    + \frac{k}{2\pi}\int_\Delta a \wedge \tilde{F}
    - \frac{k}{2\pi}\int_{\partial \Delta} a \wedge \tilde{a} 
    + \frac{p}{2\pi}\int_{\partial \Delta} a \wedge \Omega,
\end{align}
where $k$ and $p$ are integers, and the last term is the Wilson line term introduced in Section \ref{Section_6}. We can straightforwardly apply the covariant phase space formalism to \eqref{Action_4D} to compute the corresponding charges and symplectic structure: we do so in \eqref{Maxwell_Equation}--\eqref{eom_5} and \eqref{Symplectic_Potential}--\eqref{Bracket_Charges} below and show that one obtains precisely the charges that FP find in \cite{FP2018}, as well as the desired central charge. 

Before proceeding to these results, however, it is instructive to consider an alternative form of the action that yields the same equations of motion and symplectic structure, but which yields a different magnetic charge. Integrating (\ref{Action_4D}) by parts, we note that up to an exact term, the action (\ref{Action_4D}) is equivalent to the following combination of undressed fields:
\begin{align}
    \label{Action_4D_bis}
    S = \frac{1}{2e^{2}}\int_M F \wedge \star F 
    - \frac{k}{2\pi}\int_{\Delta} \tilde{A} \wedge F
    + \frac{k}{2\pi}\int_{\partial\Delta} \tilde{\phi}\wedge F
    + \frac{p}{2\pi}\int_{\partial \Delta} a \wedge \Omega,
\end{align}
where we now see that we have transformed this action into the form of our earlier low-energy action (\ref{BF1}) along with two additional terms that live on a codimension $2$ stratum.
Evidently, the third term has to be understood in the following way:
\begin{align}
\label{integral_diff_cohom}
\int_{S^{1}\times\mathbb{R}}\tilde{\phi}dA
=\sum\limits_{i=1}^{n-1}\int_{\arc{P_{i}P_{i+1}}\times\mathbb{R}}\tilde{\phi}_{i}dA
+\sum\limits_{i=1}^{n}\int_{P_{i}\times\mathbb{R}}\tilde{m}_{i\left(i+1\right)}A
\end{align}
where the $P_{i}$'s are points on $S^{1}$ and
\begin{equation}
\label{Descent_Edge_Mode}
    \tilde{\phi}_{i}-\tilde{\phi}_{i+1} = -\tilde{m}_{i\left(i+1\right)}\in 2\pi\mathbb{Z}
\end{equation}
in a neighborhood of $P_{i+1}$, which is coming directly from $D^{\left[0,0\right]}\phi = 0$). The elements $\tilde{\phi}_{i}$ and $\tilde{m}_{ij}$ transform as
\begin{align}
\label{Gauge_Transformation_Edge_Mode}
\begin{cases}
&\tilde{\phi}_{i} \longrightarrow\tilde{\phi}_{i} - \tilde{n}_{i}\\
&\tilde{m}_{ij} \longrightarrow \tilde{m}_{ij} + \tilde{n}_{i} - \tilde{n}_{j}
\end{cases}
\end{align}
that is
\begin{equation}
    \tilde{\phi}\longrightarrow \tilde{\phi} + D^{\left[0,-1\right]}\tilde{n}
\end{equation}
where
\begin{equation}
\label{D_coboundary_-1}
    D^{\left[0,-1\right]} =
    \begin{pmatrix}
    -d \\
    \check{\delta} \\
    \end{pmatrix}
\end{equation}
and 
\begin{equation}
\tilde{n} = \left(\left(\tilde{n}_{i}\right)_{i}\right)    
\end{equation}
in such a way that $D^{\left[0,0\right]}\circ D^{\left[0,-1\right]} = 0$. More generally, the sign of the $d$'s in $D^{\left[k,l\right]}$ for $k$ fixed is conventional. It simply has to flip from $l$ to $l+1$ so that $D^{\left[k,l+1\right]}\circ D^{\left[k,l\right]} = 0$. Here, it has been chosen in such a way that $D^{\left[0,0\right]}$ can be regarded as a submatrix of $D^{\left[1,0\right]}$, cf. Fig. \ref{Truncated_Complex_1} and Fig. \ref{Truncated_Complex_0} for visualizing the truncated double complexes.

An important remark is in order here: Since $\tilde{\phi}$ is defined only locally, different choices of local representatives and choices of a geometric decomposition like the $P_{i}$'s in \eqref{integral_diff_cohom} would lead to integrals that differ by $n\in 2\pi\mathbb{Z}$. Thus, only the complex exponential of those integrals (which are the natural quantum objects) is well-defined. However, for what we are interested in, i.e. the symplectic charges and bracket, this ambiguity does not matter because it is eliminated by the functional derivative.
 
The fact that (\ref{Action_4D_bis}) yields the same equations of motion and pre-symplectic form as (\ref{Action_4D}) but not the same charges can be traced to part of the so-called JKM ambiguity \cite{Kirklin_2019}: Under the transformation $L \mapsto L + dK$, we have on the boundary $S_{\partial} \mapsto S_{\partial} + \int_{\partial} K$ and the pre-symplectic potential $\theta \mapsto \theta + \delta K $ while the symplectic structure $\Omega = \delta\theta$ remains invariant.

%\textcolor{red}{Is it on purpose that we write $\frac{k}{2\pi}\int_{\partial\Delta} \tilde{\phi}\wedge F$ and not $\frac{k}{2\pi}\int_{\partial\Delta} d\tilde{\phi}\wedge A$ in \eqref{Action_4D_bis}? The latter appears more naturally when going from equation \eqref{Action_4D} to equation\eqref{Action_4D_bis} and is actually perfectly well defined, so we could skip the paragraph above. However, maybe, on the contrary, this paragraph is important. That's an occasion to explain how the integral works with local objects.}}

There are several reasons to be interested in the form of the action given in (\ref{Action_4D_bis}). First, we note that this is essentially the extended BF action discussed in \cite{geiller2019extended}, and thus on the basis of that analysis, we would expect a central charge to arise. Second, the term $\tilde{\phi}\wedge F$ that is needed to compensate for the failure of gauge invariance of $\tilde{A} \wedge F$ (in the presence of a codimension $2$ stratum) can be given the following interpretation: It is a $2$-dimensional BF action term that can function as a duality wall operator. Thus, what we would expect is that such a wall operator term in the action serves to dualize the electric Wilson loop $\int_{\gamma} a$. We now see that this is confirmed by varying the action \eqref{Action_4D} to obtain the following equations of motion:
\begin{align}
    d\star dA &= 0 \mbox{ on } M,\label{Maxwell_Equation}\\
    \star dA &= e^{2}\frac{k}{2\pi} d\tilde{A} \mbox{ on } \Delta,\label{Duality_Equation}\\
    \frac{k}{2\pi}\tilde{a} &= \frac{p}{2\pi}\Omega \mbox{ on } \partial\Delta,\label{eom_Omega}\\
    \star dA &= 0 \mbox{ on } \partial M\setminus\Delta,\label{eom_4}\\
    F &= 0 \mbox{ on } \Delta (\mbox{ and } \partial\Delta)\label{eom_5}.
\end{align}
Equation \eqref{Maxwell_Equation} is nothing but the Maxwell equation, whereas \eqref{Duality_Equation} is the duality equation that enforces the codimension $1$ EM duality. Equation \eqref{eom_Omega} may look surprising. Indeed, $\tilde{a} = d\tilde{\phi}-\tilde{A}$ and $\tilde{A}$ is globally defined on $\Delta$ which is contractible. Hence, by the Poincaré lemma, $\tilde{A}$ has to be exact, which might give the impression that $\Omega$ itself is exact, which would contradict the fact that $\Omega$ is a nontrivial de Rham cocycle. The reason this impression is false is that $\tilde{\phi}$ is \textbf{not} globally defined, although $d\tilde{\phi}$ is. Hence, 
\begin{equation}
    \label{Chern_number_liker}
    \int_{S^{1}_{\varepsilon}\left(p_{t}\right)}\tilde{a}
    =\int_{S^{1}_{\varepsilon}\left(p_{t}\right)}d\tilde{\phi}-\tilde{A}
    =\int_{S^{1}_{\varepsilon}\left(p_{t}\right)}d\tilde{\phi}
\end{equation}
by Stokes theorem, since $\tilde{A}$ is truly exact. Note that the last integral in equation \eqref{Chern_number_liker} formally looks like a first Chern number, except that in this case it is computed from a connection over a $\left(-1\right)$-gerbe instead of a usual $\mathrm{U}\!\left(1\right)$-connection. In fact, we can understand this quantity in terms of winding number, and, through the long exact sequence in cohomology for the pair $\left(\Delta,\partial\Delta\right)$, in terms of the Chern number associated with the relative Chern class of the bundle over $\Delta$ with a chosen trivialization over $\partial\Delta$. The computation of this remaining integral can be treated in a similar way as \eqref{integral_diff_cohom}. Using the same notation as before, we have:
\begin{align}
\label{integral_dphi}
\nonumber
\int_{S^{1}_{\varepsilon}\left(p_{t}\right)}d\tilde{\phi}
=&\sum\limits_{i=1}^{n-1}\int_{\arc{P_{i}P_{i+1}}}d\tilde{\phi}_{i}\\ 
\nonumber
=&\sum\limits_{i=1}^{n-1}\left(\tilde{\phi}_{i}\left(P_{i+1}\right)-\tilde{\phi}_{i}\left(P_{i}\right)\right)\\
\nonumber
=&\sum\limits_{i=1}^{n-1}\left(\tilde{\phi}_{i}\left(P_{i+1}\right)-\tilde{\phi}_{i+1}\left(P_{i+1}\right)\right)\\
\int_{S^{1}_{\varepsilon}\left(p_{t}\right)}d\tilde{\phi}
=&-\sum\limits_{i=1}^{n-1}\tilde{m}_{i\left(i+1\right)}\in 2\pi\mathbb{Z}.
\end{align}
Hence, we can truly write:
\begin{equation}
    \label{integral_eom}
    \frac{k}{2\pi}\int_{S^{1}_{\varepsilon}\left(p_{t}\right)}\tilde{a}
    =\frac{p}{2\pi}\int_{S^{1}_{\varepsilon}\left(p_{t}\right)}\Omega
\end{equation}
%Strictly speaking, \eqref{integral_dphi} is ill-defined, as its value depends on the choice of the local representatives $\tilde{\phi}_{i}$ and also the choice of the decomposition of $S^{1}_{\varepsilon}\left(p_{t}\right)$, i.e. the $P_{i}$'s.
Note that since the $\tilde{m}_{i}$'s define a class in $H^{1}\left(\partial\Delta;\mathbb{Z}\right)$ that indexes the isomorphism class of the $\left(-1\right)$-gerbes on $\partial\Delta$ over which the connection $\tilde{\phi}$ is defined, it is clear that $\Omega$ determines this class on-shell through the equation of motion. It is thus essential to have a nontrivial gerbe, otherwise $\tilde{\phi}$ would be globally defined and \eqref{integral_dphi} would simply be zero, making the equation of motion \eqref{eom_Omega} impossible to satisfy in general.

We now proceed to computing the presymplectic potential from the action \eqref{Action_4D}:
\begin{align}
\nonumber
    \theta 
    =& \frac{1}{e^{2}}\int_{\Sigma}\delta A\wedge\star dA
    + \frac{k}{2\pi}\int_{\Delta\cap\Sigma}\delta\phi\wedge d\tilde{A}
    - \frac{k}{2\pi}\int_{\partial\Delta\cap\Sigma}\delta\phi\wedge\tilde{a}\\
    &+ \frac{p}{2\pi}\int_{\partial\Delta\cap\Sigma}\delta\phi\wedge\Omega
    - \frac{k}{2\pi}\int_{\Delta\cap\Sigma}a\wedge\delta\tilde{A}
    + \frac{k}{2\pi}\int_{\partial\Delta\cap\Sigma}a\wedge\delta\tilde{\phi}
\end{align}
where $\Sigma$ is a Cauchy surface, that is, a spacelike submanifold of $M$ corresponding to $B^{3}\times \left\lbrace t_{0}\right\rbrace$ where a tubular/disk neighborhood of a puncture on $\partial B^{3} = S^{2}$ has been removed so that $\partial\Sigma\cong B^{2}\times\left\lbrace t_{0}\right\rbrace$. Note that $\delta\tilde{\phi}$ is globally defined. Indeed, if we apply $\delta$ (functional differential) to \eqref{Descent_Edge_Mode}, we obtain
\begin{equation}
    \delta\tilde{\phi}_{i}-\delta\tilde{\phi}_{i+1} = \delta\tilde{m}_{i\left(i+1\right)}
\end{equation}
and we claim 
\begin{equation}
\delta\tilde{m}_{i\left(i+1\right)}=0, 
\end{equation}
which is equivalent to saying that, when we vary a connection $\tilde{\phi}$ over a gerbe whose isomorphism class is indexed by $\tilde{m}$, we vary it among the connections over the gerbes of the same isomorphism class. The intuitive idea behind this claim is the following: The space of (piecewise) smooth $\mathbb{Z}$-valued functions is made of disconnected sheets (indexed by $\tilde{m}$), so one cannot continuously distort an element out of its sheet; hence we have
\begin{equation}
    \delta\tilde{\phi}_{i} = \delta\tilde{\phi}_{i+1} =:\delta\phi.
\end{equation}
On-shell, the presymplectic potential simplifies to:
\begin{align}
\label{Symplectic_Potential}
    \theta 
    = \frac{1}{e^{2}}\int_{\Sigma}\delta A\wedge\star dA
    + \frac{k}{2\pi}\int_{\Delta\cap\Sigma}\delta\phi\wedge d\tilde{A}
    - \frac{k}{2\pi}\int_{\Delta\cap\Sigma}a\wedge\delta\tilde{A}
    + \frac{k}{2\pi}\int_{\partial\Delta\cap\Sigma}a\wedge\delta\tilde{\phi}
\end{align}

We want to consider now a vector field that generates the (on-shell) physical symmetry of the edge modes. For that, we dualize $\delta\phi$ and $\delta\tilde{\phi}$ by $\dfrac{\delta}{\delta\phi}$ and $\dfrac{\delta}{\delta\tilde{\phi}}$ respectively, which are globally defined, since $\delta\phi$ and $\delta\tilde{\phi}$ are. The dualization works as:
\begin{equation}
    \dfrac{\delta}{\delta\phi\left(x\right)}\left(\delta\phi\left(y\right)\right) = \delta\left(x-y\right)
\end{equation}
and identically for $\delta\tilde{\phi}$ and $\dfrac{\delta}{\delta\tilde{\phi}}$.

From the homotopy pullback, the vector field that generates the symmetry of $\phi$ is globally defined, so it can be written as
\begin{equation}
    \delta_{\alpha}:=\int_{\partial M}\alpha\dfrac{\delta}{\delta\phi}
\end{equation}
(recall that $\alpha$ is of the same type as $\varepsilon$), whereas the vector field that generates the symmetry of $\tilde{\phi}$ is only locally defined because of the coefficients $\tilde{\alpha_{i}}$ (which are of the same type as $\tilde{\varepsilon}$):
\begin{equation}
    \delta_{\tilde{\alpha}}
    :=\left(\delta_{\tilde{\alpha}_{i}}\right)
    :=\left(\int_{\partial M}\tilde{\alpha}_{i}\dfrac{\delta}{\delta\tilde{\phi}}\right).
\end{equation}

Following \cite{FP2018}, we can use these symmetries to calculate the electric charge:
\begin{align}
    \nonumber
    Q^{E}
    =& \theta\left(\delta_{\alpha}\right)\\
    \nonumber
    =& \frac{k}{2\pi}\int_{x\in\Delta\cap\Sigma}
    \delta\phi\left(x\right)\left(\int_{y\in\partial M}
    \alpha\left(y\right)\dfrac{\delta}{\delta\phi\left(y\right)}\right)
    \wedge d\tilde{A}\left(x\right)\\
    \nonumber
    =& \frac{k}{2\pi}\int_{x\in\Delta\cap\Sigma}
    \left(\int_{y\in\partial M}
    \alpha\left(y\right)\delta\phi\left(x\right)\left(\dfrac{\delta}{\delta\phi\left(y\right)}\right)\right)
    \wedge d\tilde{A}\left(x\right)\\
    \nonumber
    =& \frac{k}{2\pi}\int_{x\in\Delta\cap\Sigma}
    \left(\int_{y\in\partial M}\alpha\left(y\right)\delta\left(x-y\right)\right)
    \wedge d\tilde{A}\left(x\right)\\
    \nonumber
    =& \frac{k}{2\pi}\int_{x\in\Delta\cap\Sigma}
    \alpha\left(x\right)
    \wedge d\tilde{A}\left(x\right)\\
    Q^{E}
    =& \frac{k}{2\pi}\int_{\Delta\cap\Sigma}\alpha\wedge d\tilde{A} 
\end{align}
and in the same manner the collection of magnetic charges in each open set $U_{i}$:
\begin{align}
    Q^{M}
    = \left(Q^{M}_{i}\right)
    =\left(\theta\left(\delta_{\tilde{\alpha}_{i}}\right)\right)
    =\left(\frac{k}{2\pi}\int_{\partial\Delta\cap\Sigma}a\wedge\tilde{\alpha}_{i}\right). 
\end{align}

The symplectic structure is calculated (on-shell) as:
\begin{align}
\nonumber
    \Omega 
    =& -\delta\theta\\
    =& \frac{1}{e^{2}}\int_{\Sigma}\delta A\wedge\star d\delta A
    + \frac{k}{2\pi}\int_{\Delta\cap\Sigma}\delta\phi\wedge d\delta\tilde{A}
    + \frac{k}{2\pi}\int_{\Delta\cap\Sigma}\delta a\wedge\delta\tilde{A}
    - \frac{k}{2\pi}\int_{\partial\Delta\cap\Sigma}\delta a\wedge\delta\tilde{\phi}
\end{align}

from which we finally obtain the bracket of charges:
\begin{align}
    \label{Bracket_Charges}
    \left\lbrace Q^{E},Q^{M}\right\rbrace 
    = \Omega\left(\delta_{\alpha},\delta_{\tilde{\alpha}}\right)
    = -\frac{k}{2\pi}\int_{\partial\Delta\cap\Sigma}\alpha\wedge d\tilde{\alpha}_{i}.
\end{align}
This is precisely the central charge that was suggested in \cite{FP2018}, for which we have now provided a systematic justification. 
We note that, strictly speaking, what we have here is a collection of local brackets $\left\lbrace Q^{E},Q^{M}_{i}\right\rbrace$ which turn out to be globally well-defined since they involve $d\tilde{\alpha}_{i}$, which is a globally well-defined $1$-form. 

%\textcolor{blue}{End with discusion of Gukov-Witten operators and extension of bundle.}

%ARGH, actually we can't compute charges now, obviously, because we have to go to codimension 2 to see these charges. 
%Thought: we shouldn't be allowed to just add exact terms to the action if these disappear by Stokes (of course they would modify the potential but that's pretty arbitrary).
%So actually it's pointless to compute charges or add edge modes until we motivate a codimension 2 stratum -- that's what we do in the next section. 
%Now notice: upon introducing our codimension 2 stratum, our BF theory is in almost the same form as Geiller's BF theory + edge modes. So we should also get a non-vanishing central charge like him.

%Note that, in this work, the Hodge stars, the wedge products and the integrals involve in general globally and well-defined objects. Recall in particular that the exterior derivative of a connection over a gerbe being a well-defined global form (the curvature of the connection), while a connection over a gerbe, in general, is just a collection of local forms satisfying some descent equations. The only exception is the case of \eqref{Action_4D_bis} which is well defined only up to an integer, this integer depending on the good cover and the choice of representatives in the differential cohomology classes of the fields.

\section{The scalar-2-form duality case}
\label{Section_8}

The situation we studied in the previous section can be generalized to the case where $A$ is a global $n$-form over $M$ (that can be interpreted as a connection over a trivial $\left(n-1\right)$-gerbe over $M$), $\phi$ is a global $\left(n-1\right)$-form over $\partial M$ (that can be interpreted as a connection over a trivial $\left(n-2\right)$-gerbe over $\partial M$), $\tilde{A}$ is a global $m$-form over $\Delta$ (that can be interpreted as a connection over a trivial $\left(m-1\right)$-gerbe over $\Delta$), and $\tilde{\phi}$ is a connection over a non-trivial $\left(m-2\right)$-gerbe over $\partial\Delta$. In this case, $M = B^{m+n+1}\times\mathbb{R}$, so $\partial M = S^{m+n}\times\mathbb{R}$, $\Delta = B^{m+n}\times\mathbb{R}\subset\partial M$ and $\partial\Delta = S^{m+n-1}\times\mathbb{R}$.

One can generalize the scenario even further to manifolds $M = N\times\mathbb{R}$, $N$ being any manifold with boundary, with nontrivial gerbes for $A$ ($A$ being then a more general element in $Z^{n}_{\mathrm{DB}}\left(M\right)$), $\phi$ ($\phi$ being then a more general element in $Z^{n-1}_{\mathrm{DB}}\left(\partial M\right)$), $\tilde{A}$ ($\tilde{A}$ being then a more general element in $Z^{m}_{\mathrm{DB}}\left(\Delta\right)$) and $\tilde{\phi}$ ($\tilde{\phi}$ being then a more general element in $Z^{m-1}_{\mathrm{DB}}\left(\partial M\right)$). Recall that $p$-gerbes over a manifold $X$ are classified up to isomorphism by $H^{p+2}\left(X;\mathbb{Z}\right)$, a $0$-gerbe being nothing but a usual $\mathrm{U}\!\left(1\right)$-bundle.

We now apply this framework to the much simpler case considered in \cite{Campiglia_2019}, where $n=2$ and $m=0$ on $B^{3}\times\mathbb{R}$. We have $\Delta = \partial M = S^{2}\times \mathbb{R}$, so $\partial \Delta = \varnothing$; in other words, we have no (regularized) puncture, and therefore no edge mode $\tilde{\phi}$, i.e. no connection over a $\left(-2\right)$-gerbe. In \cite{Campiglia_2019}, the authors formulated a scalar-$2$-form duality, where one of the goals was to compute the charge of the $2$-form field $B$. While the authors introduce magnetic monopoles for $B$, they note that another way of describing the scenario would be to keep the field approach for massive particles and only introduce the duality at the boundary. The latter (which is of course the approach of \cite{FP2018}) is the description that we will now provide, albeit in the case of a finite (as opposed to asymptotic) boundary.

Using the notations of \cite{Campiglia_2019}, the action we will now consider is:
\begin{equation}
    S = -\frac{1}{2}\int_M H \wedge \star H + \int_{\partial M} b \wedge d\psi
\end{equation}
where $b = d\omega - B$ ($a = d\phi - A$ with the previous notations) is the dressed field $B$ ($A$ with the previous notations) and $H = dB$ ($F=dA$ with the previous notations). The field $B$ is assumed to be a globally defined $2$-form (more abstractly, this is a connection over a $1$-gerbe, that is, in general, a collection of $2$-forms defined in the open sets of a good cover satisfying some descent equations, but here the gerbe is trivial and we choose a representative of the connection that is globally defined) and the edge mode $\omega$ is a globally defined $1$-form. The pair $\left(B,\omega\right)$ transforms as $\left(B,\omega\right)\longrightarrow \left(B + d\beta,\omega+\beta\right)$ where $\beta$ is a globally defined $1$-form. This makes $b$ gauge invariant. We note that the transformation $\omega\longrightarrow \omega + \gamma$ (where $\gamma$ is a globally defined $1$-form) alone (i.e. no associated transformation $B \longrightarrow B + d\gamma$) is an on-shell symmetry of the action. Finally, the scalar field $\psi$ ($\tilde{A}$ with the previous notations) is also assumed to be globally defined. It transforms as $\psi\longrightarrow \psi + dn$ where $n$ is an integer and $d$ represents the operator that formally injects the set of integers into the set of functions. We note that $\psi$ is here a connection over a $\left(-1\right)$-gerbe over $\Delta$ and since $\left(-1\right)$-gerbes are classified by $H^{1}\left(\Delta;\mathbb{Z}\right) = 0$, there is only one isomorphism class of $\left(-1\right)$-gerbes over $\Delta$, namely the trivial class. Moreover, since $\Delta = \partial M$, so $\partial\Delta = \varnothing$, there is no dual edge mode ($\tilde{\phi}$ with the previous notations).

The equations of motion are:
\begin{align}
    d\star dB =& 0 \mbox{ on } M,\\
    \star dB =& -d\psi \mbox{ on } \partial M,\\
    dB =& 0 \mbox{ on } \partial M. 
\end{align}

The symplectic potential is then:
\begin{align}
    \theta =
    -\frac{1}{2}\int_{\Sigma}\delta B\wedge\star dB
    +\int_{\partial M\cap\Sigma}\delta\omega\wedge d\psi
    +\int_{\partial M\cap\Sigma}b\wedge \delta\psi
\end{align}
and we can compute the charge associated with the edge mode $\omega$:
\begin{align}
    Q=\theta\left(\delta_{\alpha}\right) = \int_{\partial M\cap\Sigma}\alpha\wedge d\psi
\end{align}
which is, on-shell,
\begin{align}
    Q=\theta\left(\delta_{\alpha}\right) = -\int_{\partial M\cap\Sigma}\alpha\wedge\star H
\end{align}
This form of the charge looks almost the same as the $2$-form charge computed in \cite{Campiglia_2019}, but with one crucial difference: The symmetry used in \cite{Campiglia_2019} to compute the charge is a generalized global symmetry of $B$, whereas here it is a physical symmetry of the edge mode $\omega$ that comes from the homotopy pullback.

\begin{acknowledgments}
We thank Laurent Freidel, Daniele Pranzetti, Alexander Schenkel, Stephan Stolz, Pavel Mnev, Konstantin Wernli, Donald Youmans and Frank Thuillier for their helpful remarks on the present work and our fruitful discussions. 
\end{acknowledgments}

\appendix
\section{Detailed computations of the equations of motion}
\label{Appendix_A}

We consider the following action
\begin{align}
    S
    =\frac{1}{2e^{2}}\int_{M}dA\wedge\star dA 
    + \frac{k}{2\pi}\int_{\Delta}a\wedge d\tilde{A}
    - \frac{k}{2\pi}\int_{\partial\Delta}a\wedge\tilde{a}
    + \frac{p}{2\pi}\int_{\partial\Delta}a\wedge\Omega
\end{align}
where $a=d\phi - A$ and $\tilde{a}=d\tilde{\phi} - \tilde{A}$.

Let's vary $S$ with respect to $A$:
\begin{align}
\nonumber
    \left.\frac{d}{dt}\right|_{t=0}&S\left(A+t\delta A\right)\\
\nonumber    
    =&\frac{1}{e^{2}}\int_{M}d\delta A\wedge\star dA 
    - \frac{k}{2\pi}\int_{\Delta}\delta A\wedge d\tilde{A}
    + \frac{k}{2\pi}\int_{\partial\Delta}\delta A\wedge\tilde{a}
    - \frac{p}{2\pi}\int_{\partial\Delta}a\wedge\Omega\\
    \nonumber
    =&\frac{1}{e^{2}}\int_{M}\delta A\wedge d\star dA 
    + \frac{1}{e^{2}}\int_{M}d\left(\delta A\wedge\star dA\right)
    - \frac{k}{2\pi}\int_{\Delta}\delta A\wedge d\tilde{A}\\
    &+ \frac{k}{2\pi}\int_{\partial\Delta}\delta A\wedge\tilde{a}
    - \frac{p}{2\pi}\int_{\partial\Delta}a\wedge\Omega
\end{align}
leading to the four first equations of motion \eqref{Maxwell_Equation} -- \eqref{eom_4}.

Now, let's vary $S$ with respect to $\phi$:
\begin{align}
\nonumber
    \left.\frac{d}{dt}\right|_{t=0}&S\left(\phi+t\delta \phi\right)\\
\nonumber
    =&\frac{k}{2\pi}\int_{\Delta}d\delta\phi\wedge d\tilde{A}
    - \frac{k}{2\pi}\int_{\partial\Delta}d\delta\phi\wedge\tilde{a}
    + \frac{p}{2\pi}\int_{\partial\Delta}d\delta\phi\wedge\Omega\\
\nonumber
    =&\frac{k}{2\pi}\int_{\Delta}d\left(\delta\phi\wedge d\tilde{A}\right)
    - \frac{k}{2\pi}\int_{\partial\Delta}d\left(\delta\phi\wedge\tilde{a}\right)
    + \frac{k}{2\pi}\int_{\partial\Delta}\delta\phi\wedge d\tilde{A}\\
    &+ \frac{p}{2\pi}\int_{\partial\Delta}d\left(\delta\phi\wedge\Omega\right)
\end{align}
leading to 
\begin{equation}
    d\tilde{A} = 0 \mbox{ on } \partial\Delta
\end{equation}
which is actually weaker than \eqref{eom_Omega}, as it is obtained by differentiating \eqref{eom_Omega}, taking into account the fact that $\Omega$ is closed.

Now, let's vary $S$ with respect to $\tilde{A}$:
\begin{align}
\nonumber
    \left.\frac{d}{dt}\right|_{t=0}S\left(\tilde{A} + t\delta\tilde{A}\right)
    =&\frac{k}{2\pi}\int_{\Delta}a\wedge d\delta\tilde{A}
    + \frac{k}{2\pi}\int_{\partial\Delta}a\wedge\delta\tilde{A}\\
    =&-\frac{k}{2\pi}\int_{\Delta}dA\wedge\delta\tilde{A}
    -\frac{k}{2\pi}\int_{\Delta}d\left(a\wedge\delta\tilde{A}\right)
    + \frac{k}{2\pi}\int_{\partial\Delta}a\wedge\delta\tilde{A}
\end{align}
from which we get our last equation of motion \eqref{eom_5}.

Finally, let's vary $S$ with respect to $\tilde{\phi}$:
\begin{align}
\nonumber
    \left.\frac{d}{dt}\right|_{t=0}S\left(\tilde{\phi} + t\delta\tilde{\phi}\right)
    =&-\frac{k}{2\pi}\int_{\partial\Delta}a\wedge d\delta\tilde{\phi}\\
    =&\frac{k}{2\pi}\int_{\partial\Delta}dA\wedge\delta\tilde{\phi}
    +\frac{k}{2\pi}\int_{\partial\Delta}d\left(a\wedge\delta\tilde{\phi}\right)
\end{align}
which leads to a weaker version of \eqref{eom_5} (the same equation but on $\partial\Delta$ only).

\section{Explicit verification of the gauge invariance of the presymplectic potential on-shell.}
\label{Appendix_B}

First, we have:
\begin{align}
\nonumber
    \theta\left(A+d\varepsilon,\phi + \varepsilon\right)
    - \theta\left(A,\phi\right)
    =& \frac{1}{e^{2}}\int_{\Sigma}d\delta\varepsilon\wedge\star dA
    + \frac{k}{2\pi}\int_{\Delta\cap\Sigma}\delta\varepsilon\wedge d\tilde{A}\\
\nonumber
    &- \frac{k}{2\pi}\int_{\partial\Delta\cap\Sigma}\delta\varepsilon\wedge\tilde{a}+ \frac{p}{2\pi}\int_{\partial\Delta\cap\Sigma}\delta\varepsilon\wedge\Omega\\
\nonumber
    =& - \frac{1}{e^{2}}\int_{\Sigma}d\delta\varepsilon\wedge d\star dA
    + \frac{1}{e^{2}}\int_{\Sigma}d\left(\delta\varepsilon\wedge\star dA\right)
    + \frac{k}{2\pi}\int_{\Delta\cap\Sigma}\delta\varepsilon\wedge d\tilde{A}\\
\nonumber
    &- \frac{k}{2\pi}\int_{\partial\Delta\cap\Sigma}\delta\varepsilon\wedge\tilde{a}
    + \frac{p}{2\pi}\int_{\partial\Delta\cap\Sigma}\delta\varepsilon\wedge\Omega\\
    \theta\left(A+d\varepsilon,\phi + \varepsilon\right)
    - \theta\left(A,\phi\right)
    =&0
\end{align}
on-shell since $\partial\Sigma = - \Delta\cap\Sigma$ (Indeed, both $\partial\Sigma$ and $\Delta\cap\Sigma$ are homeomorphic to $B^{2}\times\left\lbrace t_{0}\right\rbrace$ but $\partial\Sigma$ is reached through $\Sigma$ and $\Delta\cap\Sigma$ is reached through $\Delta$ so they have opposite orientation.) Otherwise:
\begin{align}
\nonumber
    \theta\left(\tilde{A} + d\tilde{\varepsilon}, \tilde{\phi} + \tilde{\varepsilon}\right)
    - \theta\left(\tilde{A},\tilde{\phi}\right)   
    &= - \frac{k}{2\pi}\int_{\Delta\cap\Sigma}a\wedge d\delta\tilde{\varepsilon}
    + \frac{k}{2\pi}\int_{\partial\Delta\cap\Sigma}a\wedge\delta\tilde{\varepsilon}\\
\nonumber
    &= \frac{k}{2\pi}\int_{\Delta\cap\Sigma}dA\wedge\delta\tilde{\varepsilon}
    + \frac{k}{2\pi}\int_{\Delta\cap\Sigma}d\left(a\wedge\delta\tilde{\varepsilon}\right)
    + \frac{k}{2\pi}\int_{\partial\Delta\cap\Sigma}a\wedge\delta\tilde{\varepsilon}\\
    \theta\left(\tilde{A} + d\tilde{\varepsilon}, \tilde{\phi} + \tilde{\varepsilon}\right)
    - \theta\left(\tilde{A},\tilde{\phi}\right)
    &=0
\end{align}
on-shell.

\section{The DB double complex}
\label{Appendix_C}

We are interested here in two distinct DB complexes, represented on Fig. \ref{Truncated_Complex_1} and Fig. \ref{Truncated_Complex_0} below, which can be regarded as \v{C}ech-de Rham double complexes with two special constraints: 
\begin{enumerate}
    \item We impose $\Omega^{-1}\left(U_{i_{0}i_{1}\hdots i_{n}}\right) = \mathbb{Z}^{N_{n+1}}$ where $N$ is the number of nonempty  intersections of $\left(n+1\right)$ open sets of the open cover, that is, we associate an integer to each intersection $U_{i_{0}i_{1}\hdots i_{n}} = U_{i_{0}}\cap U_{i_{1}}\cap\hdots\cap U_{i_{n}}$. The map 
    \begin{equation}
    d:\Omega^{-1}\left(U_{i_{0}i_{1}\hdots i_{n}}\right) = \mathbb{Z}^{N_{n+1}} \longrightarrow\Omega^{0}\left(U_{i_{0}i_{1}\hdots i_{n}}\right)= C^{\infty}\left(U_{i_{0}i_{1}\hdots i_{n}},\mathbb{R}\right)    
    \end{equation}
    is the canonical injection.
    \item We impose a truncation at level $k$ ($k=1$ for the DB complex on Fig. \ref{Truncated_Complex_1} and $k=0$ for the DB complex on Fig. \ref{Truncated_Complex_0}), which means that $\Omega^{k}\left(U_{i_{0}i_{1}\hdots i_{n}}\right)$ is mapped to $0$ with the zero map, instead of $\Omega^{k+1}\left(U_{i_{0}i_{1}\hdots i_{n}}\right)$ with the de Rham differential $d$.
\end{enumerate}
The differential $D^{\left[k,l\right]}$ maps, for $k$ fixed, the $l$-th diagonal of the double complex (truncated at level $k$) onto the $\left(l+1\right)$-th ($l\geq -1$). The sign of the $d$'s in $D^{\left[k,l\right]}$ for $k$ fixed is conventional. It simply has to flip from $l$ to $\left(l+1\right)$ so that $D^{\left[k,l+1\right]}\circ D^{\left[k,l\right]} = 0$.

\begin{figure}[ht!]
    \centering
    \includegraphics[width=\linewidth]{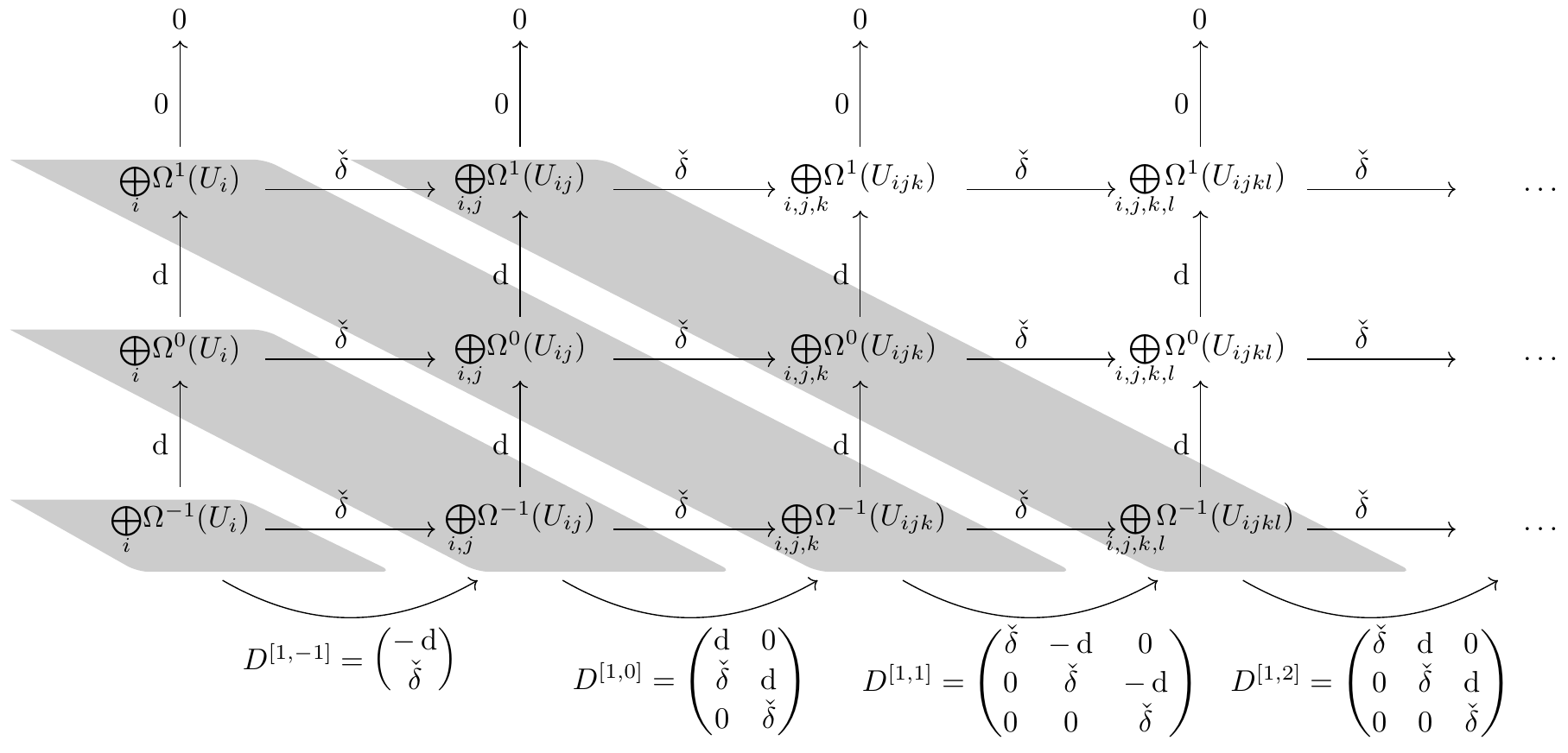}
    \caption{Complex truncated at form degree $1$}
\label{Truncated_Complex_1}
\end{figure}

\begin{figure}
    \centering
    \includegraphics[width=\linewidth]{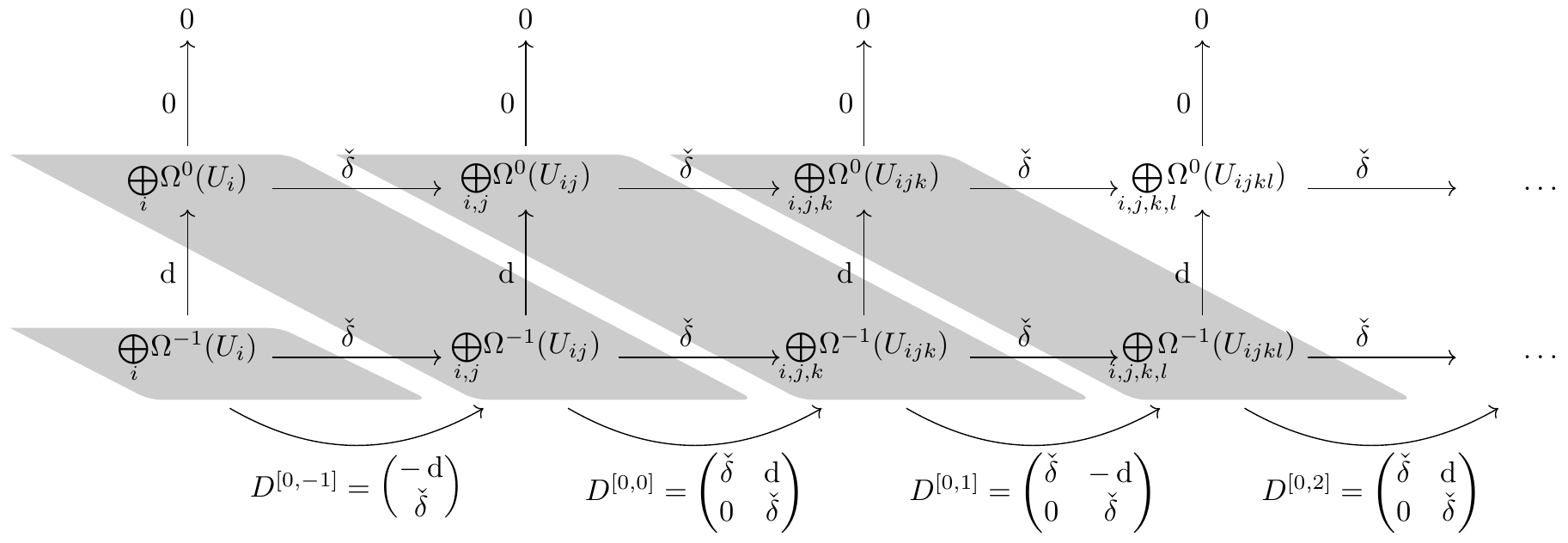}
    \caption{Complex truncated at form degree $0$}
\label{Truncated_Complex_0}
\end{figure}

\newpage

\section{Connections over an $n$-gerbe}
\label{Appendix_D}

For $n\geq -2$, $n$-gerbes over a manifold $Y$ are classified up to isomorphism by $H^{n+2}\left(Y;\mathbb{Z}\right)$, or said differently, an $n$-gerbe over a manifold $Y$ is a geometric representative of a class in $H^{n+2}\left(Y;\mathbb{Z}\right)$, and two isomorphic $n$-gerbes represent the same class in $H^{n+2}\left(Y;\mathbb{Z}\right)$. Hence, a $0$-gerbe over $Y$ is nothing but a usual $\mathrm{U}\!\left(1\right)$-bundle over $Y$. In the mathematics literature, $1$-gerbes are often called simply ``gerbes'' and aroused special interest. 

We are not really interested here in $n$-gerbes properly speaking, but rather in connections over such a structure, which we describe now. In the following, let $Y$ be a closed finite dimensional manifold provided with a good cover, that is, a collection of open sets covering $Y$ such that any intersection of these open sets is either empty or contractible.

Consider a DB $1$-cochain in the DB complex truncated at form degree $1$, that is, an element
\begin{equation}
A = \left(\left(A_{i}\right)_{i},\left(\Lambda_{ij}\right)_{ij},\left(n_{ijk}\right)_{ijk}\right)
    \in\bigoplus_{i}\Omega^{1}\!\left(U_{i}\right)
    \oplus\bigoplus_{i,j}\Omega^{0}\!\left(U_{ij}\right)
    \oplus\bigoplus_{i,j,k}\Omega^{-1}\!\left(U_{ijk}\right)
\end{equation}
(a collection of three collections: First, a collection of $1$-forms in the open sets $U_{i}$, then a collection of functions in the double intersections $U_{ij} = U_{i}\cap U_{j}$, and finally a collection of integers in the triple intersections $U_{ijk} = U_{i}\cap U_{j}\cap U_{k}$). Saying that $A$ is a connection over a $\mathrm{U}\!\left(1\right)$-bundle (or $0$-gerbe) is equivalent to saying, as we have seen earlier, that
\begin{equation}
D^{\left[1,1\right]}A=0 
\end{equation}
where 
\begin{equation}
\begin{tabular}{c}
$\bigoplus\limits_{i}\Omega^{1}\!\left(U_{i}\right)
\oplus\bigoplus\limits_{i,j}\Omega^{0}\!\left(U_{ij}\right)
\oplus\bigoplus\limits_{i,j,k}\Omega^{-1}\!\left(U_{ijk}\right)$\\
$\hspace{-1cm}D^{\left[1,1\right]}\Bigg\downarrow$\\
$\bigoplus\limits_{i,j}\Omega^{1}\!\left(U_{ij}\right)
\oplus\bigoplus\limits_{i,j,k}\Omega^{0}\!\left(U_{ijk}\right)
\oplus\bigoplus\limits_{i,j,k,l}\Omega^{-1}\!\left(U_{ijkl}\right)$
\end{tabular}
\end{equation}

This generalizes straightforwardly to the case of an $\left(n+1\right)$-cochain in the DB complex truncated at form degree $\left(n+1\right)$, that is, an element
\begin{align}
A &= \left(\left(A_{i_{0}}\right)_{i_{0}}, \left(A_{i_{0}i_{1}}\right)_{i_{0}i_{1}}, \hdots, \left(A_{i_{0}i_{1}\hdots i_{n+2}}\right)_{i_{0}i_{1}\hdots i_{n+2}}\right)\\
\nonumber
&\in\bigoplus_{i_{0}}\Omega^{n+1}\!\left(U_{i_{0}}\right)
\oplus\bigoplus_{i_{0},i_{1}}\Omega^{n}\!\left(U_{i_{0}i_{1}}\right)
\oplus\hdots
%\oplus\bigoplus_{i_{0},i_{1}\hdots}\Omega^{0}\!\left(U_{i_{0}i_{1}\hdots i_{n+1}}\right)
\oplus\bigoplus_{i_{0},i_{1}\hdots}\Omega^{-1}\!\left(U_{i_{0}i_{1}\hdots i_{n+2}}\right).
\end{align}
(a collection of $\left(n+3\right)$ collections: First, a collection of $\left(n+1\right)$-forms in the open sets $U_{i}$, then a collection of $n$-forms in the double intersections $U_{ij} = U_{i}\cap U_{j}$ and so on, we decrease the de Rham degree by $1$ and increase the \v{C}ech degree by $1$ at each step, until we get a collection of functions in the intersections of $n+2$ open sets of the cover $U_{i_{0}i_{1}\hdots i_{n+1}}$ and finally a collection of integers in the intersections of $n+3$ open sets of the cover $U_{i_{0}i_{1}\hdots i_{n+2}}$). Saying that $A$ is a connection over an $n$-gerbe is equivalent to saying that
\begin{equation}
D^{\left[n+1,n+1\right]}A=0. 
\end{equation}
where 
\begin{equation}
\begin{tabular}{c}
$\bigoplus\limits_{i_{0}}\Omega^{n+1}\!\left(U_{i_{0}}\right)
\oplus\bigoplus\limits_{i_{0},i_{1}}\Omega^{n}\!\left(U_{i_{0}i_{1}}\right)
\oplus\hdots
%\oplus\bigoplus\limits_{i_{0},i_{1}\hdots}\Omega^{0}\!\left(U_{i_{0}i_{1}\hdots i_{n+1}}\right)
\oplus\bigoplus\limits_{i_{0},i_{1}\hdots}\Omega^{-1}\!\left(U_{i_{0}i_{1}\hdots i_{n+2}}\right)$\\
$\hspace{-1cm}D^{\left[n+1,n+1\right]}\Bigg\downarrow$\\
$\bigoplus\limits_{i_{0},i_{1}}\Omega^{n+1}\!\left(U_{i_{0}i_{1}}\right)
\oplus\bigoplus\limits_{i_{0},i_{1},i_{2}}\Omega^{n}\!\left(U_{i_{0}i_{1}i_{2}}\right)
\oplus\hdots
%\oplus\bigoplus\limits_{i_{0},i_{1},\hdots}\Omega^{0}\!\left(U_{i_{0}i_{1}\hdots i_{n+1}i_{n+2}}\right)
\oplus\bigoplus\limits_{i_{0},i_{1},\hdots}\Omega^{-1}\!\left(U_{i_{0}i_{1}\hdots i_{n+3}}\right)$
\end{tabular}
\end{equation}

Furthermore, a gauge transformation of a connection 
\begin{equation}
A = \left(\left(A_{i}\right)_{i},\left(\Lambda_{ij}\right)_{ij},\left(n_{ijk}\right)_{ijk}\right)   
\end{equation}
over a $\mathrm{U}\!\left(1\right)$-bundle is a transformation 
\begin{equation}
A\longrightarrow A + D^{\left[1,0\right]}q     
\end{equation}
where
\begin{equation}
    q = \left(\left(q_{i}\right)_{i},\left(m_{ij}\right)_{ij}\right) 
    \in\bigoplus_{i}\Omega^{0}\!\left(U_{i}\right)\oplus\bigoplus_{i,j}\Omega^{-1}\!\left(U_{ij}\right).
\end{equation}

Similarly, a gauge transformation of a connection 
\begin{equation}
A = \left(\left(A_{i_{0}}\right)_{i_{0}}, 
\left(A_{i_{0}i_{1}}\right)_{i_{0}i_{1}}, 
\hdots, 
\left(A_{i_{0}i_{1}\hdots i_{n+2}}\right)_{i_{0}i_{1}\hdots i_{n+2}}\right)    
\end{equation}
over an $n$-gerbe is a transformation
\begin{equation}
A\longrightarrow A + D^{\left[n+1,n\right]}q     
\end{equation}
where
\begin{align}
    q &= \left(\left(q_{i_{0}}\right)_{i_{0}},\left(q_{i_{0}i_{1}}\right)_{i_{0}i_{1}},\hdots,\left(q_{i_{0}i_{1}\hdots i_{n+1}}\right)_{i_{0}i_{1}\hdots i_{n+1}}\right)\\
    \nonumber
    &\in\bigoplus_{i_{0}}\Omega^{n}\!\left(U_{i_{0}}\right)
    \oplus\bigoplus_{i_{0},i_{1}}\Omega^{n-1}\!\left(U_{i_{0}i_{1}}\right)
    \oplus\hdots
%   \oplus\bigoplus_{i_{0},i_{1},\hdots,i_{n}}\Omega^{0}\!\left(U_{i_{0}i_{1}\hdots i_{n}}\right)
    \oplus\bigoplus_{i_{0},i_{1},\hdots,i_{n+1}}\Omega^{-1}\!\left(U_{i_{0}i_{1}\hdots i_{n+1}}\right).
\end{align}

Finally, the $\mathbb{Z}$-module of connections over a $\mathrm{U}\!\left(1\right)$-bundle over a manifold $Y$ modulo gauge transformations, i.e. $\slfrac{\mathrm{ker}D^{\left[1,1\right]}}{\mathrm{im}D^{\left[1,0\right]}}$, is the $\mathbb{Z}$-module of DB cohomology classes of degree $1$ denoted $H^{1}_{\mathrm{DB}}\left(Y;\mathbb{Z}\right)$, and we have the following short exact sequences:
\begin{equation}
    0
    \longrightarrow
    \slfrac{\Omega^{1}\left(Y\right)}{\Omega^{1}_{\mathbb{Z}}\left(Y\right)}
    \overset{\check{\delta}}{\longrightarrow}
    H^{1}_{\mathrm{DB}}\left(Y;\mathbb{Z}\right)    
    \overset{u}{\longrightarrow}
    H^{2}\left(Y;\mathbb{Z}\right)
    \longrightarrow
    0
\end{equation}
and
\begin{equation}
    0
    \longrightarrow
    H^{1}\left(Y;\slfrac{\mathbb{R}}{\mathbb{Z}}\right)    
    \overset{v}{\longrightarrow}
    H^{1}_{\mathrm{DB}}\left(Y;\mathbb{Z}\right)    
    \overset{d}{\longrightarrow}
    \Omega^{2}_{\mathbb{Z}}\left(Y\right)
    \longrightarrow
    0
\end{equation}
where $\Omega^{k}_{\mathbb{Z}}\left(Y\right)$ stands for the closed $k$-forms over $Y$ with integral period ($k\in\left\lbrace 1,2 \right\rbrace$) and the maps $\check{\delta}$, $u$, $v$ and $d$ are explicitly given by
\begin{equation}
\check{\delta} 
\left|
\begin{array}{ccc}
        \slfrac{\Omega^{1}\left(Y\right)}{\Omega^{1}_{\mathbb{Z}}\left(Y\right)} 
        & \longrightarrow 
        & H^{1}_{\mathrm{DB}}\left(Y;\mathbb{Z}\right)\\
        \left[\omega\right] 
        & \longmapsto 
        & \left[\left(\left(\check{\delta}\omega\right)_{i}, 0, 0\right)\right],
\end{array}
\right.
\end{equation}

\begin{equation}
u 
\left|
  \begin{array}{ccc}
    H^{1}_{\mathrm{DB}}\left(Y;\mathbb{Z}\right) 
    & \longrightarrow 
    & H^{2}\left(Y;\mathbb{Z}\right)\\
    \left[\left(\left(A_{i}\right)_{i}, \left(\Lambda_{ij}\right)_{ij}, \left(n_{ijk}\right)_{ijk}\right)\right] 
    & \longmapsto 
    & \left[\left(n_{ijk}\right)_{ijk}\right],
  \end{array}
\right.
\end{equation}

\begin{equation}
v 
\left|
\begin{array}{rcl}
    H^{1}\left(\slfrac{\mathbb{R}}{\mathbb{Z}}\right) 
    & \longrightarrow 
    & H^{1}_{\mathrm{DB}}\left(\mathbb{Z}\right)\\
    \left[\left(m_{ij}\right)_{ij}\right] 
    & \longmapsto 
    & \left[\left(0, \left(dm_{ij}\right)_{ij}, \left(\check{\delta}\left(m_{ij}\right)\right)_{ijk}\right)\right],
\end{array}
\right.
\end{equation}
and finally:
\begin{equation}
d
\left|
\begin{array}{ccc}
    H^{1}_{\mathrm{DB}}\left(Y;\mathbb{Z}\right) & \longrightarrow & \Omega^{2}_{\mathbb{Z}} \\
    \left[\left(\left(A_{i}\right)_{i}, \left(\Lambda_{ij}\right)_{ij}, \left(n_{ijk}\right)_{ijk}\right)\right] 
    & \longmapsto 
    & F = \left(dA_{i}\right)_{i}
\end{array}
\right.
\end{equation}
(by definition, $\left(\check{\delta}\left(m_{ij}\right)\right)_{ijk}\in\mathbb{Z}^{N_{3}}$ and the curvature $F$ is a globally well-defined object).

Likewise, the $\mathbb{Z}$-module of connections over an $n$-gerbe over a manifold $Y$ modulo gauge transformations, i.e. $\slfrac{\mathrm{ker}D^{\left[n+1,n+1\right]}}{\mathrm{im}D^{\left[n+1,n\right]}}$, is the $\mathbb{Z}$-module of DB cohomology classes of degree $\left(n+1\right)$ denoted $H^{n+1}_{\mathrm{DB}}\left(Y;\mathbb{Z}\right)$, and we have the short exact sequences
\begin{equation}
    0
    \longrightarrow
    \slfrac{\Omega^{n+1}\left(Y\right)}{\Omega^{n+1}_{\mathbb{Z}}\left(Y\right)}
    \overset{\check{\delta}}{\longrightarrow}
    H^{n+1}_{\mathrm{DB}}\left(Y;\mathbb{Z}\right)    
    \overset{u}{\longrightarrow}
    H^{n+2}\left(Y;\mathbb{Z}\right)
    \longrightarrow
    0
\end{equation}
and
\begin{equation}
    0
    \longrightarrow
    H^{n+1}\left(Y;\slfrac{\mathbb{R}}{\mathbb{Z}}\right)    
    \overset{v}{\longrightarrow}
    H^{n+1}_{\mathrm{DB}}\left(Y;\mathbb{Z}\right)    
    \overset{d}{\longrightarrow}
    \Omega^{n+2}_{\mathbb{Z}}\left(Y\right)
    \longrightarrow
    0
\end{equation}
where the maps $\check{\delta}$, $u$, $v$ and $d$ are exactly the same as above.

Let's emphasize here that the convention considered so far for degrees is the DB convention. Concretely, the DB degree of a DB cohomology class is the form degree of the first family of components of the representatives of the class. There is a $\left(+1\right)$ shift from the DB convention to the differential cohomology convention, e.g. a DB $1$-cocycle is a differential $2$-cocycle. For a physicist, the DB convention is more relevant as the DB degree of a class matches with the degree of the field considered. But for a mathematician, the differential cohomology convention makes more sense as the cup product we can define is truly graded commutative.

\bibliographystyle{unsrt}
\bibliography{duality}

\end{document}